\documentclass[%
 reprint,
superscriptaddress,
nofootinbib,
 amsmath,amssymb,
 aps, prd
]{revtex4-1}
\usepackage{graphicx}
\usepackage{dcolumn}
\usepackage{bm}
\usepackage{hyperref}
\usepackage{bm}
\usepackage{subfigure}
\usepackage{comment}
\usepackage{xcolor}
\usepackage{multirow}
\usepackage[mathlines]{lineno}
\usepackage{footmisc}

\usepackage{ifthen} 
\newboolean{pdflatex}
\setboolean{pdflatex}{true} 

\newboolean{articletitles}
\setboolean{articletitles}{true} 

\newboolean{uprightparticles}
\setboolean{uprightparticles}{false} 

\newboolean{inbibliography}
\setboolean{inbibliography}{false} 

\newboolean{wordcount}
\setboolean{wordcount}{false} 

\usepackage{xspace}
\makeatletter
\gdef\@ptsize{0} 
\makeatother
\makeatletter
\usepackage{suffix}
\let\latex@section\section
\WithSuffix\def\section*{\secdef\my@section{\latex@section*}}
\def\my@section[#1]#2{}
\makeatother

\newcommand {\bdstlnu} {\overline{B} \rightarrow D^\ast \ell^- \overline{\nu}_\ell}
\newcommand {\bdlnu} {\overline{B} \rightarrow D \ell^- \overline{\nu}_\ell}

\newcommand {\barnuell} {\overline{\nu}_\ell}

\def\GF   {G_F} 
\def\BB      {\ensuremath{B\overline{B}}\xspace}
\def\qsq     {{\ensuremath{q^2}}\xspace}

\def\ctl     {\ensuremath{\cos{\theta_\ell}}\xspace}

\def\thetal  {\ensuremath{\theta_\ell}\xspace}

\usepackage{relsize}
\def\babar{\mbox{\slshape B\kern-0.1em{\smaller A}\kern-0.1em B\kern-0.1em{\smaller A\kern-0.2em R}}}

\newcommand {\ups} {\Upsilon(4S)}
\newcommand {\eex} {E_{\mbox{\scriptsize extra}}}

\newcommand {\btag} {B_{\mbox{\scriptsize tag}}}
\newcommand {\bsig} {\overline{B}_{\mbox{\scriptsize sig}}}

\def\DeltaE     {\mbox{$\Delta E$}\xspace}

\def\mes        {\mbox{$m_{\rm ES}$}\xspace}
\newcommand {\bsigb} {\overline{B}_{\mbox{\scriptsize sig}}}

\newcommand {\vcb} {V_{cb}}
\newcommand {\vub} {V_{ub}}

\def\ellm       {{\ensuremath{\ell^-}}\xspace}
\def\deriv {\ensuremath{\mathrm{d}}}

\usepackage{xspace} 
\usepackage{upgreek}

\def\MagUp {\mbox{\em Mag\kern -0.05em Up}\xspace}

\ifthenelse{\boolean{uprightparticles}}%
{

 \def\Ppi         {\ensuremath{\uppi}\xspace}

 \def\PDelta      {\ensuremath{\Delta}\xspace}                 
 \def\PXi      {\ensuremath{\Xi}\xspace}                 
 \def\PLambda      {\ensuremath{\Lambda}\xspace}                 
 \def\PSigma      {\ensuremath{\Sigma}\xspace}                 
 \def\POmega      {\ensuremath{\Omega}\xspace}                 
 \def\PUpsilon      {\ensuremath{\Upsilon}\xspace}                 
 
 \def\PB      {\ensuremath{\mathrm{B}}\xspace}                 
                  
 \def\PD      {\ensuremath{\mathrm{D}}\xspace}

 \def\PK      {\ensuremath{\mathrm{K}}\xspace}

 \def\Pb      {\ensuremath{\mathrm{b}}\xspace}                 
 \def\Pc      {\ensuremath{\mathrm{c}}\xspace}

 \def\Pi      {\ensuremath{\mathrm{i}}\xspace}

 \def\Pu      {\ensuremath{\mathrm{u}}\xspace}

}
{

 \def\Ppi         {\ensuremath{\pi}\xspace}

 \mathchardef\PDelta="7101
 \mathchardef\PXi="7104
 \mathchardef\PLambda="7103
 \mathchardef\PSigma="7106
 \mathchardef\POmega="710A
 \mathchardef\PUpsilon="7107
                  
 \def\PB      {\ensuremath{B}\xspace}                 
                  
 \def\PD      {\ensuremath{D}\xspace}

 \def\PK      {\ensuremath{K}\xspace}

 \def\Pb      {\ensuremath{b}\xspace}                 
 \def\Pc      {\ensuremath{c}\xspace}

 \def\Pi      {\ensuremath{i}\xspace}

 \def\Pu      {\ensuremath{u}\xspace}

}

\makeatletter
\ifcase \@ptsize \relax
  \newcommand{\miniscule}{\@setfontsize\miniscule{4}{5}}
\or
  \newcommand{\miniscule}{\@setfontsize\miniscule{5}{6}}
\or
  \newcommand{\miniscule}{\@setfontsize\miniscule{5}{6}}
\fi
\makeatother

\DeclareRobustCommand{\optbar}[1]{\shortstack{{\miniscule (\rule[.5ex]{1.25em}{.18mm})}
  \\ [-.7ex] $#1$}}

\def\ellm       {{\ensuremath{\ell^-}}\xspace}

\def\uquark    {{\ensuremath{\Pu}}\xspace}

\def\cquark    {{\ensuremath{\Pc}}\xspace}

\def\bquark    {{\ensuremath{\Pb}}\xspace}

\def\pion   {{\ensuremath{\Ppi}}\xspace}
\def\piz    {{\ensuremath{\pion^0}}\xspace}
\def\pip    {{\ensuremath{\pion^+}}\xspace}
\def\pim    {{\ensuremath{\pion^-}}\xspace}

  \def\Kbar    {{\kern 0.2em\overline{\kern -0.2em \PK}{}}\xspace}

\def\KorKbar    {\kern 0.18em\optbar{\kern -0.18em K}{}\xspace}

  \def\Dbar    {{\kern 0.2em\overline{\kern -0.2em \PD}{}}\xspace}
\def\D       {{\ensuremath{\PD}}\xspace}

\def\DorDbar    {\kern 0.18em\optbar{\kern -0.18em D}{}\xspace}
\def\Dz      {{\ensuremath{\D^0}}\xspace}

\def\Dp      {{\ensuremath{\D^+}}\xspace}

\def\B       {{\ensuremath{\PB}}\xspace}
\def\Bbar    {{\ensuremath{\kern 0.18em\overline{\kern -0.18em \PB}{}}}\xspace}

\def\BorBbar    {\kern 0.18em\optbar{\kern -0.18em B}{}\xspace}
\def\Bz      {{\ensuremath{\B^0}}\xspace}

  \def\Y#1S{\ensuremath{\PUpsilon{(#1S)}}\xspace}

\def\Lbar        {{\ensuremath{\kern 0.1em\overline{\kern -0.1em\PLambda}}}\xspace}
\def\LorLbar    {\kern 0.18em\optbar{\kern -0.18em \PLambda}{}\xspace}

\def\to                 {\ensuremath{\rightarrow}\xspace}

\def\qsq       {{\ensuremath{q^2}}\xspace}

\def\Vub  {{\ensuremath{V_{\uquark\bquark}}}\xspace}
\def\Vcb  {{\ensuremath{V_{\cquark\bquark}}}\xspace}

\def\AT#1     {\ensuremath{A_{\mathrm{T}}^{#1}}\xspace}           

\def\ctl       {\ensuremath{\cos{\theta_\ell}}\xspace}

\def\C#1      {\ensuremath{\mathcal{C}_{#1}}\xspace}                       
\def\Cp#1     {\ensuremath{\mathcal{C}_{#1}^{'}}\xspace}                    
\def\Ceff#1   {\ensuremath{\mathcal{C}_{#1}^{\mathrm{(eff)}}}\xspace}        
\def\Cpeff#1  {\ensuremath{\mathcal{C}_{#1}^{'\mathrm{(eff)}}}\xspace}       
\def\Ope#1    {\ensuremath{\mathcal{O}_{#1}}\xspace}                       
\def\Opep#1   {\ensuremath{\mathcal{O}_{#1}^{'}}\xspace}                    

\newcommand{\tev}{\ifthenelse{\boolean{inbibliography}}{\ensuremath{~T\kern -0.05em eV}}{\ensuremath{\mathrm{\,Te\kern -0.1em V}}}\xspace}
\newcommand{\gev}{\ensuremath{\mathrm{\,Ge\kern -0.1em V}}\xspace}
\newcommand{\mev}{\ensuremath{\mathrm{\,Me\kern -0.1em V}}\xspace}
\newcommand{\kev}{\ensuremath{\mathrm{\,ke\kern -0.1em V}}\xspace}
\newcommand{\ev}{\ensuremath{\mathrm{\,e\kern -0.1em V}}\xspace}
\newcommand{\gevc}{\ensuremath{{\mathrm{\,Ge\kern -0.1em V\!/}c}}\xspace}
\newcommand{\mevc}{\ensuremath{{\mathrm{\,Me\kern -0.1em V\!/}c}}\xspace}
\newcommand{\gevcc}{\ensuremath{{\mathrm{\,Ge\kern -0.1em V\!/}c^2}}\xspace}
\newcommand{\gevgevcccc}{\ensuremath{{\mathrm{\,Ge\kern -0.1em V^2\!/}c^4}}\xspace}
\newcommand{\mevcc}{\ensuremath{{\mathrm{\,Me\kern -0.1em V\!/}c^2}}\xspace}

\def\invfb   {\ensuremath{\mbox{\,fb}^{-1}}\xspace}

\def\deriv {\ensuremath{\mathrm{d}}}

\def\gsim{{~\raise.15em\hbox{$>$}\kern-.85em
          \lower.35em\hbox{$\sim$}~}\xspace}
\def\lsim{{~\raise.15em\hbox{$<$}\kern-.85em
          \lower.35em\hbox{$\sim$}~}\xspace}

\def\tell1  {TELL1\xspace}
\def\ukl1   {UKL1\xspace}

\begin{document}

\onecolumngrid
\small
\begin{flushleft}
\babar-PUB-23/02 \\ 
SLAC-PUB-17730
\end{flushleft}
\normalsize

\title{Model-independent extraction of form factors and $|\Vcb|$ in $\overline{B} \rightarrow D \ell^- \overline{\nu}_\ell$ with hadronic tagging at \babar}

\author{J.~P.~Lees}
\author{V.~Poireau}
\author{V.~Tisserand}
\author{E.~Grauges}
\author{A.~Palano}
\author{G.~Eigen}
\author{D.~N.~Brown}
\author{Yu.~G.~Kolomensky}
\author{M.~Fritsch}
\author{H.~Koch}
% ...
\collaboration{The \babar\ Collaboration}
\noaffiliation

\begin{abstract}
\noindent Using the entire $\babar$ $\ups$ data set, the first two-dimensional unbinned angular analysis of the semileptonic decay $\bdlnu$ is performed, employing hadronic reconstruction of the tag-side $B$ meson from $\ups\to \BB$. Here, $\ell$ denotes the light charged leptons $e$ and $\mu$. A novel data-driven signal-background separation procedure with minimal dependence on simulation is developed. This procedure preserves all multi-dimensional correlations present in the data. The expected $\sin^2\thetal$ dependence of the differential decay rate in the Standard Model is demonstrated, where $\thetal$ is the lepton helicity angle. Including input from the latest lattice QCD calculations and previously available experimental data, the underlying form factors are extracted using both model-independent (BGL) and dependent (CLN) methods. Comparisons with lattice calulations show flavor SU(3) symmetry to be a good approximation in the $B_{(s)}\to D_{(s)}$ sector. Using the BGL results, the CKM matrix element $|\Vcb|=(41.09\pm 1.16)\times 10^{-3}$ and the Standard Model prediction of the lepton-flavor universality violation variable $\mathcal{R}(D)=0.300\pm 0.004$, are extracted. The value of $|\Vcb|$ from $\bdlnu$ tends to be higher than that extracted using $\bdstlnu$. The Standard Model $\mathcal{R}(D)$ calculation is at a $1.97\sigma$ tension with the latest HFLAV experimental average. 

\end{abstract}

\let\oldmaketitle\maketitle
\renewcommand\maketitle{{\bfseries\boldmath\oldmaketitle}}
\maketitle

\clearpage 

\pagebreak
\section{Introduction}

The $\bdlnu$ decay is one of the better understood semileptonic (SL) $B$ meson decays. The Cabibbo-favored nature of the underlying tree-level ${b \to c W^{\ast -}}$ transition leads to large branching fractions. The spin-0 nature of the $B$ and $D$ mesons dictates that the $c$-quark hadronization is described by a single form factor (FF) for the massless lepton case~\cite{Richman:1995wm,spd-paper}. Due to these inherent simplifications, the $\bdlnu$ decay is suitable for extracting the Cabibbo-Kobayashi-Maskawa (CKM)~\cite{Cabibbo:1963yz,Kobayashi:1973fv} matrix element $|\Vcb|$. In the so-called unitarity triangle of the Standard Model (SM), the length of the side opposite to the angle $\beta$ is proportional to the ratio $|\vub|/|\vcb|$. Given that $\sin 2\beta$ is measured via loop-level processes to better than $2\%$~\cite{CKMfitter2005} relative uncertainty, precise tree-level determinations of $|\Vub|$ and $|\Vcb|$ are important to test the overall consistency of the SM picture of weak interactions. However, there has been a persistent tension~\cite{Gambino:2019sif} at the level of 3 standard deviations, in both $|\Vub|$ and $|\Vcb|$, between measurements involving inclusive and exclusive final states. Following a previous study for the $\bdstlnu$ vector meson case~\cite{Dey:2019bgc}, this article deepens our understanding of this tension and the underlying FFs in the $b\to c$ sector for the pseudoscalar meson case.

In the differential decay rate of the exclusive $\bdlnu$ decay~\footnote{The inclusion of charge-conjugate decay modes is implied and natural units with $\hbar = c = 1$ are used throughout this article.}, the overall normalization is proportional to the square of the product of $|\vcb|$ and the value of a single underlying FF at the zero-recoil point, where the daughter $D$ meson is at rest in the parent $B$ meson rest frame ($w=1$ in Eq.~\ref{eqn:w_def}). However, at this zero-recoil point, the decay rate vanishes because of vanishing available phase-space, and measuring the FF shape near the zero-recoil point becomes experimentally challenging. The statistical uncertainties in this region form the dominant contribution to the uncertainty in extrapolating the FF shape to the zero-recoil point. Historically, the extrapolation has utilized theoretical expectations from heavy-quark effective theory (HQET), although the problem has been alleviated to some degree, thanks to availability of lattice QCD calculations close to the zero-recoil point in the $\overline{B}\to D$ sector~\cite{Lattice:2015rga,Na:2015kha}.

Using the entire \babar\ $\ups$ data set, we analyze the process  $e^+ e^- \to \ups \to \btag \bsigb$, where $\bsigb\to D \ellm\barnuell$, and $\btag$ is a fully reconstructed hadronic decay. Many aspects of this analysis are analogous to the recent \babar\ angular analysis of $\bdstlnu$~\cite{Dey:2019bgc}. The large data set allows for a final reconstructed $\bdlnu$ data sample with sufficient statistical precision, despite the hadronic tagging efficiency being small ($\mathcal{O}(10^{-3})$ or less). A novel event-wise signal-background separation technique is employed, preserving correlations among the different kinematic variables. Furthermore, the angular analysis employs unbinned maximum likelihood fits that avoid information loss due to binning, present in binned $\chi^2$ fits. Detector acceptance effects are handled using angular analysis techniques for exclusive $B$ meson decays~\cite{Dey:2019bgc,spd-paper}.

Several previous measurements exist for the branching fractions and the FFs in the $\bdlnu$ decay~\cite{aleph_1997,cleo_1999,belle_2002,babar_2009_global,Aubert:2009ac,Glattauer:2015teq}. In this article, updated measurements of the FF shapes are provided, and the expected $\sin^2 \thetal$ dependence of the full differential decay rate is demonstrated, where $\thetal$ is the $\ellm$ polar angle in the $W^\ast$ helicity frame. This angular dependence results  from the left-handed nature of the charged weak current of the semileptonic $W^{\ast-} \to \ell^- \barnuell$ decay in combination with the pseudoscalar nature of $B$ and $D$ mesons~\cite{Richman:1995wm,spd-paper}.
Note that this dependence is insulated from any new-physics contribution that might enter on the hadronic side $b \to c$ transition. Demonstrating the $\sin^2 \thetal$ dependence, thus establishes the reliability of the missing neutrino reconstruction as well as the signal-background separation technique.

Two functional forms of the FF parameterization are employed: first, a variant of the model-independent Boyd-Grinstein-Lebed~\cite{Boyd:1995sq} (BGL) $z$-expansion method adopted in Refs.~\cite{Glattauer:2015teq,Lattice:2015rga}; second, the more model-dependent form due to Caprini-Lellouch-Neubert~\cite{Caprini:1997mu} (CLN), which incorporates HQET and QCD sum rules. In addition to data from \babar, available data from Belle~\cite{Glattauer:2015teq} and results from lattice QCD~\cite{Lattice:2015rga,Na:2015kha} calculations are incorporated. Lattice QCD results typically cover a limited kinematic region close to the zero-recoil point. Recently, the HPQCD Collaboration has published lattice QCD FFs covering the entire kinematic region in the di-lepton mass squared, $\qsq$, for the related $B_s\to D_s^{(\ast)}$~\cite{McLean:2019qcx,Harrison:2021tol} modes. Under the assumption of flavor-SU(3) relations, spectator-quark effects can be ignored and the $B_s\to D_s^{(\ast)}$ FFs can be connected to the $B\to D^{(\ast)}$ FFs. It is important to validate flavor-SU(3) symmetry assumptions in the simpler case for $B_{(s)}\to D_{(s)}$, which can provide insight for the more complicated $B_{(s)}\to D^\ast_{(s)}$ case. 

\section{Differential decay rate and form factors}
\label{sec:diff_rate}

Ignoring scalar and tensor interaction terms, which would arise from new-physics contributions, the amplitude for $\bdlnu$ derives solely from the vector interaction term~\cite{spd-paper}
\begin{linenomath}
\begin{align}
\label{eqn:ff_def}
\langle D |\bar{c} \gamma_\mu b|\Bbar\rangle_V  &= f_+(\qsq) \left((p_B + p_D)_\mu  - \frac{ (p_B + p_D)\cdot q }{\qsq} q_\mu \right) \nonumber \\
 & \hspace{1cm} + f_0(\qsq) \frac{ (p_B + p_D)\cdot q }{\qsq} q_\mu,
\end{align}
\end{linenomath}
where $p_B$ and $p_D$ are the 4-momenta of the $B$ and $D$ mesons, respectively, and $q=p_B - p_D$ is the 4-momentum of the recoiling $(\ell^- \barnuell)$ system. The vector and scalar FFs are $f_+(\qsq)$ and $f_0(\qsq)$, respectively, corresponding to specific spin states of the $\overline{B}D$ system. In HQET, the FFs in Eq.~\ref{eqn:ff_def} are written in the form~\cite{Lattice:2015rga}
\begin{linenomath}
\begin{align}
\label{eqn:ff_def_hqet}
\frac{\langle D |\bar{c} \gamma_\mu b|\Bbar\rangle_V}{\sqrt{m_B m_D}}= h_+(w) (v+v')_\mu + h_-(w) (v-v')_\mu,
\end{align}
\end{linenomath}
where $v$ and $v'$ are the 4-velocities of $B$ and $D$ mesons, respectively, and $w=v\cdot v'$ is the relativistic $\gamma$ factor of the daughter $D$ meson in the mother $B$ meson's rest frame,
\begin{linenomath}
\begin{align}
\label{eqn:w_def}
w &= \frac{m_B^2 + m^2_D - \qsq}{2 m_B m_D}.
\end{align}
\end{linenomath}
The two sets of FFs are related as
\begin{linenomath}
\begin{subequations}
\label{eqn:rel_hqet_conv_ff}
\begin{align}
f_+(\qsq) &= \frac{1}{2\sqrt{r}}  \left((1+r)h_+(w) - (1-r)h_-(w)\right)\\
f_0(\qsq)&= \sqrt{r}\left(\frac{w+1}{1+r}h_+(w) - \frac{w-1}{1-r}h_-(w)\right),
\end{align}
\end{subequations}
\end{linenomath}
where $r=m_D/m_B$. This leads to the relation at the maximum recoil, $\qsq=0$ (neglecting the lepton masses), 
\begin{linenomath}
\begin{align}
\label{eqn:maxrecoilrel}
f_0(0) &= f_+(0).
\end{align}
\end{linenomath}

For the light (approximately massless) leptons $\ell = \{e,\mu\}$, ignoring tensor and higher order interactions, the $\bdlnu$ amplitude depends on a single FF $f_+(\qsq)$. The differential rate can be written as~\cite{spd-paper}
\begin{linenomath}
\begin{align}
\frac{\deriv\Gamma}{\deriv\qsq \deriv\ctl} = \frac{\GF^2 |\Vcb|^2 \eta^2_{\rm EW}}{32 \pi^3}k^3 |f_+(\qsq)|^2 \sin^2\thetal,
\label{eqn:rate_formula}
\end{align}
\end{linenomath}
where $k=m_D\sqrt{w^2-1}$ is the magnitude of the $D$ meson 3-momentum in the $B$ meson rest frame. Here, $\eta_{\rm EW}=1.0066$~\cite{Sirlin:1981ie} denotes leading electroweak corrections and $G_F$ is the Fermi decay constant. The FF $f_+(w)$ is sometimes also referred to as $\mathcal{G}(w)$, with the connection
\begin{linenomath}
\begin{align}
\mathcal{G}(w)^2 = \frac{4r}{(1+r)^2} f_+(w)^2.
\end{align}
\end{linenomath}

\subsection{The BGL form}
\label{sec:bgl}

The BGL~\cite{Boyd:1995sq} form employs an expansion in the variable
\begin{linenomath}
\label{eq:bgl_z_expan}
\begin{align}
z(w) = (\sqrt{w+1} - \sqrt{2})/(\sqrt{w+1} + \sqrt{2}),
\end{align}
\end{linenomath}
which is small in the physical kinematic region. The FFs are written as 
\begin{linenomath}
\begin{equation}
 f_i(z) = \frac{1}{P_i(z) \phi_i(z)} \sum\limits_{n=0}^{N} a^i_{n} z^n, \quad i\in\{+,0\},
\label{eqn:BGL}
\end{equation}
\end{linenomath}
where $P_i(z)$ are the Blaschke factors that remove contributions of bound state $B_c^{(*)}$ poles, and $\phi_i(z)$ are non-perturbative outer functions. The coefficients $a^i_n$ are free parameters and $N$ is the order at which the series is truncated. Following Refs.~\cite{Lattice:2015rga,Glattauer:2015teq}, the parameterizations adopted are
\begin{linenomath}
\begin{subequations}
\label{eqn:phi_defs}
\begin{align}
P_i(z)=&1 \\
\phi_+(z) =&\;  1.1213 (1+z)^2 (1-z)^{1/2} \nonumber \\
          & \times [(1+r)(1-z)+ 2\sqrt{r}(1+z)]^{-5}, \\
\phi_0(z) =&\;  0.5299 (1+z)(1-z)^{3/2}  \nonumber \\
          & \times    [(1+r)(1-z)+ 2\sqrt{r}(1+z)]^{-4}. 
\end{align}
\end{subequations}
\end{linenomath}
The coefficients $a^i_{n}$ in Eq.~\ref{eqn:BGL} satisfy the unitarity condition $\sum_n |a^i_{n}|^2\leq 1$. 

\subsection{The CLN form}
\label{sec:cln}

Taking into account QCD dispersion relations and based on HQET, the CLN~\cite{Caprini:1997mu} parameterization is 
\begin{linenomath}
\begin{align} 
\mathcal{G}(w) =&\; \mathcal{G}(1)(1 -8 \rho^2_D z(w) \nonumber \\
&+ (51\rho^2_D -10) z(w)^2 - (252 \rho^2_D -84) z(w)^3),
\label{eq:cln}
\end{align}
\end{linenomath}
where $z$ is the same as in the BGL expansion. This is the form that has conventionally been used in previous $\bdlnu$ analyses~\cite{Aubert:2009ac,Lees:2013uzd,belle_2002}, convenient because of the compact form of the parameterization in terms of just two variables: the normalization $\mathcal{G}(1)$, and the slope, $\rho^2_D$. It is to be noted that the relation between the slope and curvature in Eq.~\ref{eq:cln} has been scrutinized in several updated HQET analyses, such as in Ref.~\cite{Bernlochner:2022ywh}, and found to be over-constraining.

\subsection{Semi-tauonic observables}
\label{sec:rd_def}

The differential rate given in Eq.~\ref{eqn:rate_formula} for the massless lepton case can be generalized to include effects due to non-zero lepton mass, $m_\ell$. In this case, the differential rates are~\cite{Bailey:2012jg}
\begin{linenomath}
\begin{subequations}
\begin{align}
\frac{\deriv\Gamma^+}{\deriv \qsq} =& \frac{G^2}{16 \pi^3} \left(1 - \frac{m_\ell^2}{\qsq}\right)^2 k \frac{m_\ell^2}{\qsq} \nonumber \\
& \times \left(\frac{k^2 f^2_+}{3} + \frac{(m_B^2-m^2_D)^2}{4 m_B^2} f^2_0 \right), \\
\frac{\deriv\Gamma^-}{\deriv \qsq} =& \frac{G^2}{24 \pi^3} \left(1 - \frac{m_\ell^2}{\qsq}\right)^2 k^3 f^2_+, \\
\Gamma(\qsq,m_\ell) =& \frac{\deriv\Gamma^+}{\deriv \qsq} + \frac{\deriv\Gamma^-}{\deriv \qsq},
\end{align}
\end{subequations}
\end{linenomath}
where the superscripts denote the lepton helicity in the $W^{\ast -}$ rest frame and $G =\GF|\Vcb|\eta_{\rm EW}$. The ratio $\mathcal{R}(D)$ is defined as
\begin{linenomath}
\begin{align}
\mathcal{R}(D) = \frac{\int_{m^2_\tau}^{(m_B-m_D)^2} \Gamma(\qsq,m_\tau) \deriv \qsq   \textcolor{white}{\big|_{\ell = e/\mu} }  }{ \int_{m^2_\ell}^{(m_B-m_D)^2} \Gamma(\qsq,m_\ell) \deriv \qsq\big|_{\ell = e/\mu}  }.
\end{align}
\end{linenomath}

\section{Event selection}
\label{sec:sel}

\subsection{The \babar\ detector and data set}

The data used in this analysis were collected with the \babar\ detector at the PEP-II asymmetric-energy $e^+e^-$ $B$-factory at the SLAC National Accelerator Laboratory. It operated at a center of mass (c.m.) energy of 10.58~GeV at the peak of the $\ups$ resonance, which decays almost exclusively to \BB pairs. The data sample comprises 471 million $\ups\to \BB$ events, corresponding to an integrated luminosity of 426~\invfb~\cite{BaBar:2013agn}.

Charged particles are reconstructed using a tracking system, consisting of a silicon-strip detector (SVT) and a drift chamber (DCH). Particle identification of charged tracks is performed based on their ionization energy loss in the tracking devices and by a ring-imaging Cerenkov detector (DIRC). A finely segmented CsI(Tl) calorimeter (EMC) measures the energy and position of electromagnetic showers generated by electrons and photons. The EMC is surrounded by a superconducting solenoid providing a 1.5 T magnetic field and by a segmented flux return with a hexagonal barrel section and two endcaps. The steel of the flux return is instrumented (IFR) with resistive plate chambers and limited streamer tubes to detect particles penetrating the magnet coil and steel. A detailed description of the \babar\ detector can be found in Refs.~\cite{Aubert:2001tu,TheBABAR:2013jta}.

\subsection{Simulation samples}
\label{sec:sim_samples}

To identify background components, optimize selection criteria, and correct for reconstruction and detector-related inefficiencies, a sample of simulated $B\overline{B}$ events approximately 10 times larger than the \babar\ data set is used. The decay of the pairs of neutral or charged $B$ mesons from $\ups \to \btag\bsig$ is handled in a generic fashion according to their known decay modes, using the EvtGen~\cite{Lange:2001uf} package. Simulated non-$\ups$ events corresponding to the $q\overline{q}$ continuum are also included. The $q\overline{q}$ fragmentation is performed by Jetset~\cite{Sjostrand:1993yb}, and the detector response by Geant4~\cite{GEANT4:2002zbu}. Radiative effects such as bremsstrahlung in the detector material and initial-state and final-state radiation~\cite{Barberio:1993qi} are included. This simulation sample is termed {\tt GENBB} and is generated centrally for all \babar\ analyses. The simulated events are reweighted to update the associated branching fractions and FF models to more recent values. After the reweighting, this sample is the same as employed in previous \babar\ analyses~\cite{Dey:2019bgc,Lees:2013uzd}. 

\subsection{The full hadronic reconstruction}

Full hadronic reconstruction of the $\btag$ in the process $e^+ e^- \to \ups \to \btag \bsigb$ is a powerful technique that produces a clean sample of $\bsig$ mesons with undetected neutrinos. This analysis utilizes the same tagging procedure as that in several previous \babar\ analyses~\cite{Dey:2019bgc,manuel_prl,Lees:2013uzd}. The $\btag$ candidate is reconstructed in its decay into a charm-meson seed ${S \in \{D^{(*)0},D^{(*)+},D^{(*)+}_s,J/\psi\}}$ plus a system, $Y$, of charmless light hadrons, with at most five charged and two neutral particles. The $\btag$ candidate reconstruction relies on two variables that are almost uncorrelated
\begin{linenomath}
\begin{align}
\DeltaE = E_{\mbox{\scriptsize tag}}^\ast -  \sqrt{s}/2, \label{eqn:DeltaE} \\
\mes = \sqrt{s/4 -|\vec{p}_{\mbox{\scriptsize tag}}^\ast|^2} \label{eqn:mes}
\end{align} 
\end{linenomath}
where $\sqrt{s}$ is the c.m.\ energy obtained from the precisely known energies of the colliding beams, and $E^\ast_{\mbox{\scriptsize tag}}$ and $\vec{p}^\ast_{\mbox{\scriptsize tag}}$ are the reconstructed energy and 3-momentum of the candidate $\btag$ in the c.m.\ frame. To select a clean $\bsig$ sample, $\mes > 5.27$~GeV and $|\DeltaE| < 72$~MeV are required on the tag side. 

\subsection{Signal side reconstruction}
\label{sec:dlnu_sel}

The selection requirements for the lepton and $D$ meson candidates on the signal side, for the most part, follow those in the previous \babar\ analyses~\cite{Dey:2019bgc,Lees:2013uzd}. Each $\btag$ candidate is combined with a $D$ meson and a charged lepton $\ell \in \{e,\mu\}$ such that the overall charge is zero. No additional charged tracks are allowed to be associated with the event candidate, but additional photons are allowed. The laboratory momentum of the charged lepton is required to be greater than 200~MeV and 300~MeV for electrons and muons, respectively. The $D$ meson reconstruction modes used in this analysis are tabulated in Table~\ref{table:modes}. Only the five cleanest accessible $D$ meson modes are included, as listed in Table~\ref{table:modes}. At this stage, the reconstructed invariant masses of the $D$ meson candidates are required to be within four standard deviations of the expected resolution around their nominal masses.

\begin{linenomath}
\begin{table}
   \caption{ \label{table:modes} The five $D$ meson decay modes and two leptonic modes used in this analysis with the final signal and background yields for the amplitude analysis after all selection requirements (see Sec.~\ref{sec:final_yields}). }
  \begin{center}
     \begin{tabular}{clcccc} \\ \hline \hline
     $\ellm$ &  $D$ & decay mode & mode & $N_{\rm sig}$ & $N_{\rm bkgd}$\\ \hline
     \multirow{3}{*}{$e^-$} &   \multirow{3}{*}{\Dz}  & $K^- \pip$ & 0 & $ 539$ & $ 63$ \\
       &         & $K^- \pip \piz$ & 1 &  $ 813$ & $ 196$     \\
       &          & $K^- \pip \pim \pip$ & 2 & $ 550$ & $ 82$   \\ \hline
      \multirow{2}{*}{$e^-$} &  \multirow{2}{*}{\Dp}     
                  & $K^- \pip \pip$  & 3  &   $ 721$ & $ 41$   \\
       &          & $K^- \pip \pip \piz$ & 4 &   $ 204$ & $ 120$ \\ \hline
     \multirow{3}{*}{$\mu^-$} &   \multirow{3}{*}{\Dz}  & $K^- \pip$ & 5 & $ 433$ & $ 64$  \\
       &         & $K^- \pip \piz$ & 6 &    $ 798$ & $ 221$   \\
       &          & $K^- \pip \pim \pip$ & 7 & $ 608$ & $ 84$  \\ \hline
      \multirow{2}{*}{$\mu^-$} &  \multirow{2}{*}{\Dp}     
                  & $K^- \pip \pip$  & 8  &  $ 665$ & $ 55$    \\
       &          & $K^- \pip \pip \piz$ & 9 &  $ 233$ & $ 134$  \\ \hline
        \multicolumn{4}{ r }{Total} & 5563 & 1061 \\ \hline \hline
    \end{tabular}
   \end{center}
\end{table}
\end{linenomath}

After selecting a $\BB$ candidate comprising $\btag$, $D$ and $\ell$, the overall missing 4-momentum is assigned to the undetected neutrino as
\begin{linenomath}
\begin{align}
p_\nu \equiv p_{\mbox{\scriptsize miss}} = p_{e^+e^-} - p_{\rm tag} - p_{D} - p_\ell.
\end{align}
\end{linenomath}
Thus, hadronic $B$ tagging allows for indirect detection of all final-state particles in semileptonic $B$ meson decays for the light leptons $\ell = \{e,\mu\}$, with a single missing neutrino. The discriminating variable is
\begin{linenomath}
\begin{align} 
U = E^{\ast \ast}_{\mbox{\scriptsize miss}} - |\vec{p}^{\;\ast\ast}_{\mbox{\scriptsize miss}}|,
\end{align} 
\end{linenomath}
where $E^{\ast \ast}_{\mbox{\scriptsize miss}}$ and $\vec{p}^{\;\ast\ast}_{\mbox{\scriptsize miss}}$ are respectively the neutrino energy and 3-momentum calculated in the $\bsig$ rest frame. The presence of a clear peak in $U$ allows for a signal extraction procedure where knowledge of the exact nature and composition of the background is relatively unimportant, as long as there is no background component that peaks in the signal region (see Fig.~\ref{fig:dlnu_stacked}).

For a given event candidate, the variable $\eex$ is defined as the sum of the energies of all additional good quality ($E_\gamma > 50$~MeV) photon candidates in the calorimeter, not associated with the reconstructed candidate. Candidates having $\eex > 0.8$~GeV are rejected, the criterion being intentionally kept loose. Next, a kinematic fit is performed on the entire event using the {\tt TreeFitter} algorithm~\cite{treefitter}. The fit constrains masses of the $\btag$, $\bsig$, and the $D$ mesons to their nominal values. In addition, the fit constrains the $D$ and $B$ meson decay products to originate from the appropriate vertex, allowing for the $\bsig$ non-zero flight length. The $\ups$ candidate vertex is also constrained to the primary vertex, within uncertainties. A nominal requirement is placed on the $\chi^2$-probability or confidence level (CL) from the fit to be greater than $10^{-10}$, to select only convergent fits. For events with multiple candidates after all selection requirements are applied, the candidate with the lowest value of $\eex$ is retained. Furthermore, this chosen candidate is required to also correspond to the one with the highest CL or else the event is rejected. For each selected event candidate, a second version of the kinematic fit is performed with an additional $U=0$ constraint corresponding to zero missing mass, as expected for a signal candidate with a single undetected neutrino. This additional constraint improves the resolution in the reconstructed kinematic variables, $\qsq$ and $\ctl$, for true signal events (see Sec.~\ref{sec:syst_uncer}). Therefore, after the signal-background separation has been performed, the further analysis uses the $\{\qsq,\thetal\}$ variables reconstructed from the kinematic fit including this zero missing mass constraint.

Each of the ten signal modes in Table~\ref{table:modes} has its own independent background and acceptance characteristics. Therefore, for further processing, the entire data set is divided into ten corresponding subsets that undergo independent background-subtraction and acceptance-correction procedures. The subsets are combined at the last stage of the analysis for the angular fit (see Eq.~\ref{eqn:log_likelihood_qval_wt}).

\section{Signal-background separation}
\label{sec:sig_bkgd}

\subsection{Introduction}

\begin{figure}
\begin{center}
\includegraphics[width=3.4in]{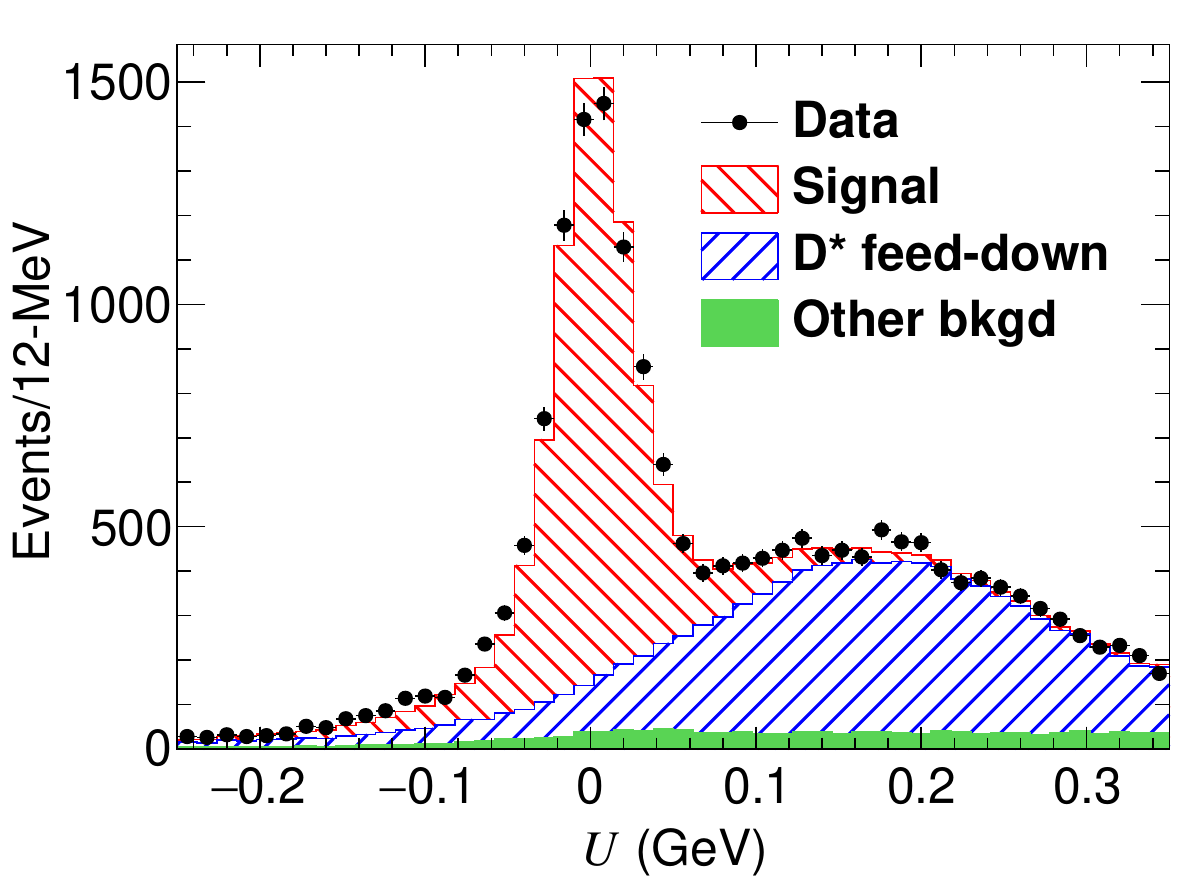}
\end{center}
\caption[]{Stacked histograms based on {\tt GENBB} simulation showing the different components of the events, after all selection requirements and integrated over the ten modes in Table~\ref{table:modes}. The data are overlaid as well but no fits in $U$ have been performed to match the data with simulation at this stage (see Sec.~\ref{sec:sig_bkgd}).}
\label{fig:dlnu_stacked}
\end{figure}

\begin{figure*}
\centering
\includegraphics[width=2.23in]{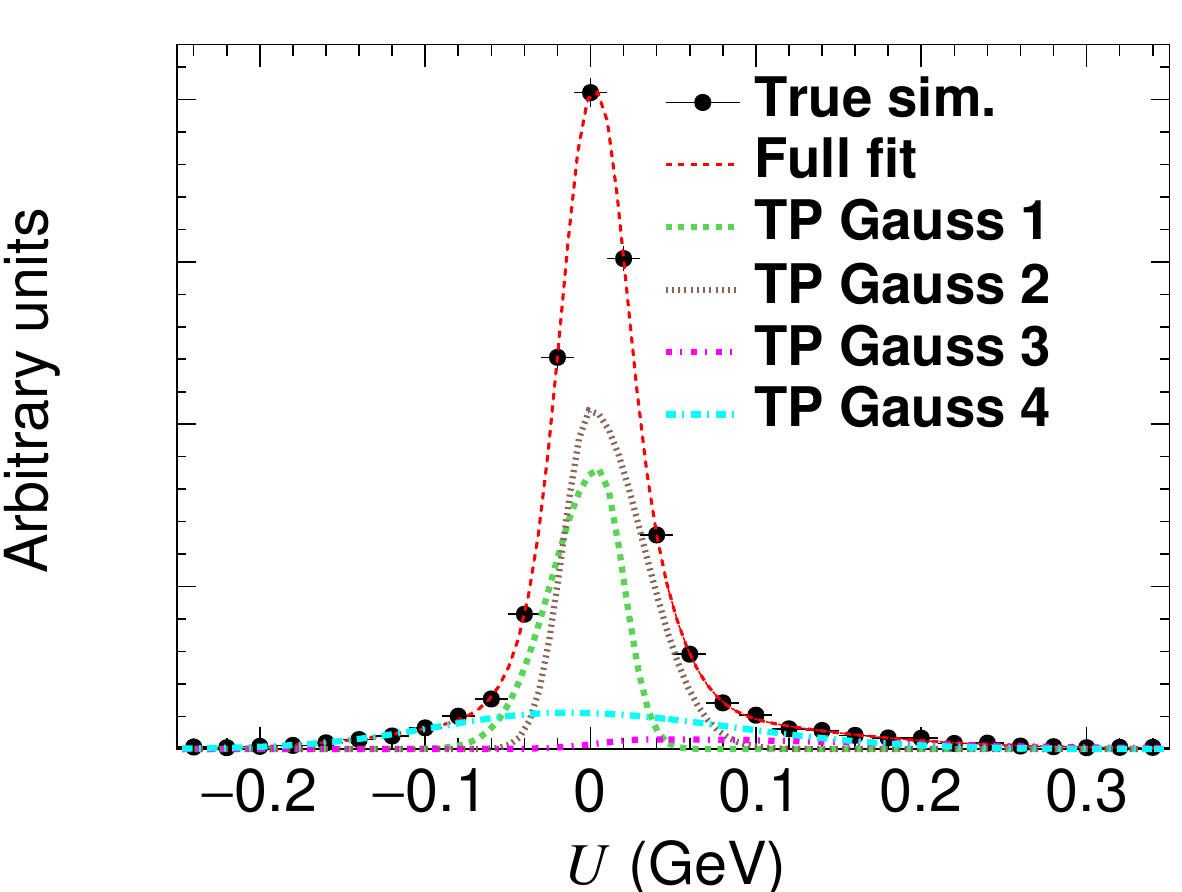}
\includegraphics[width=2.23in]{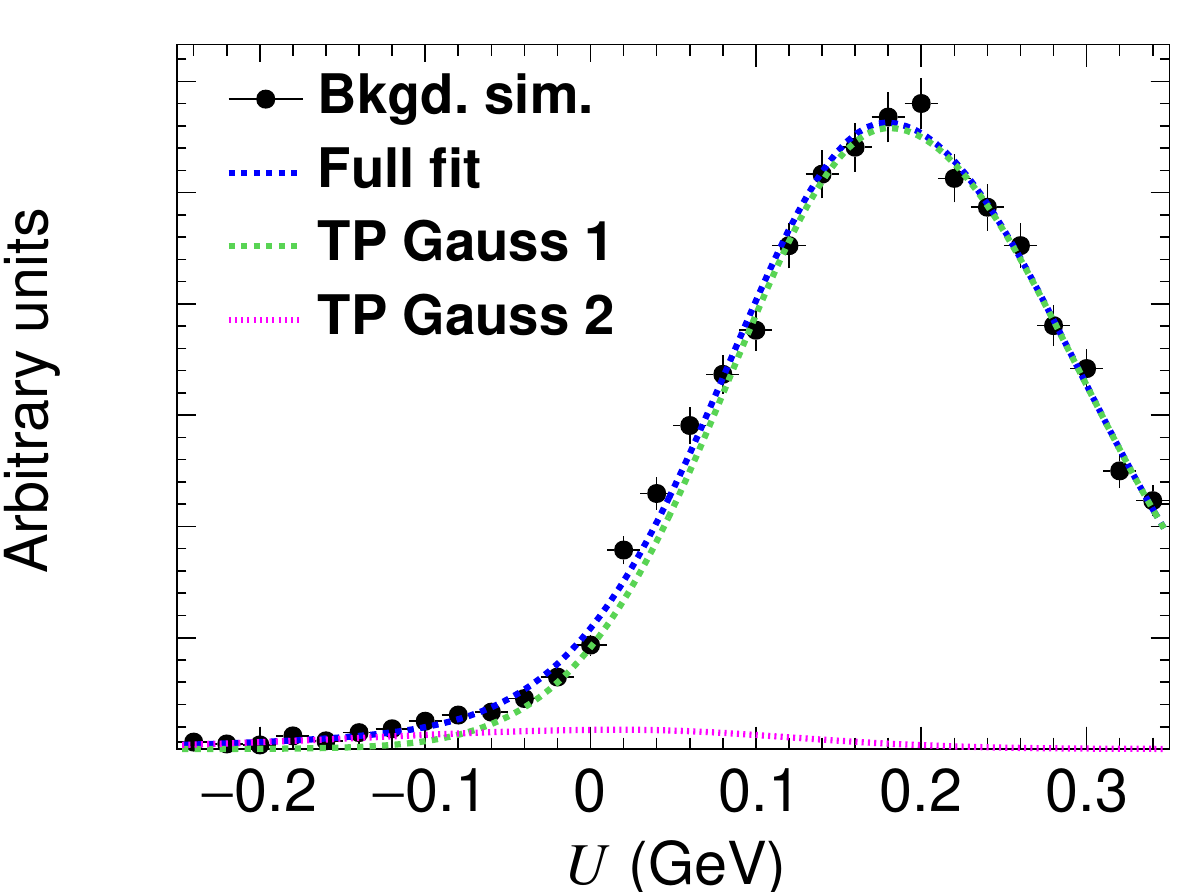}
\includegraphics[width=2.23in]{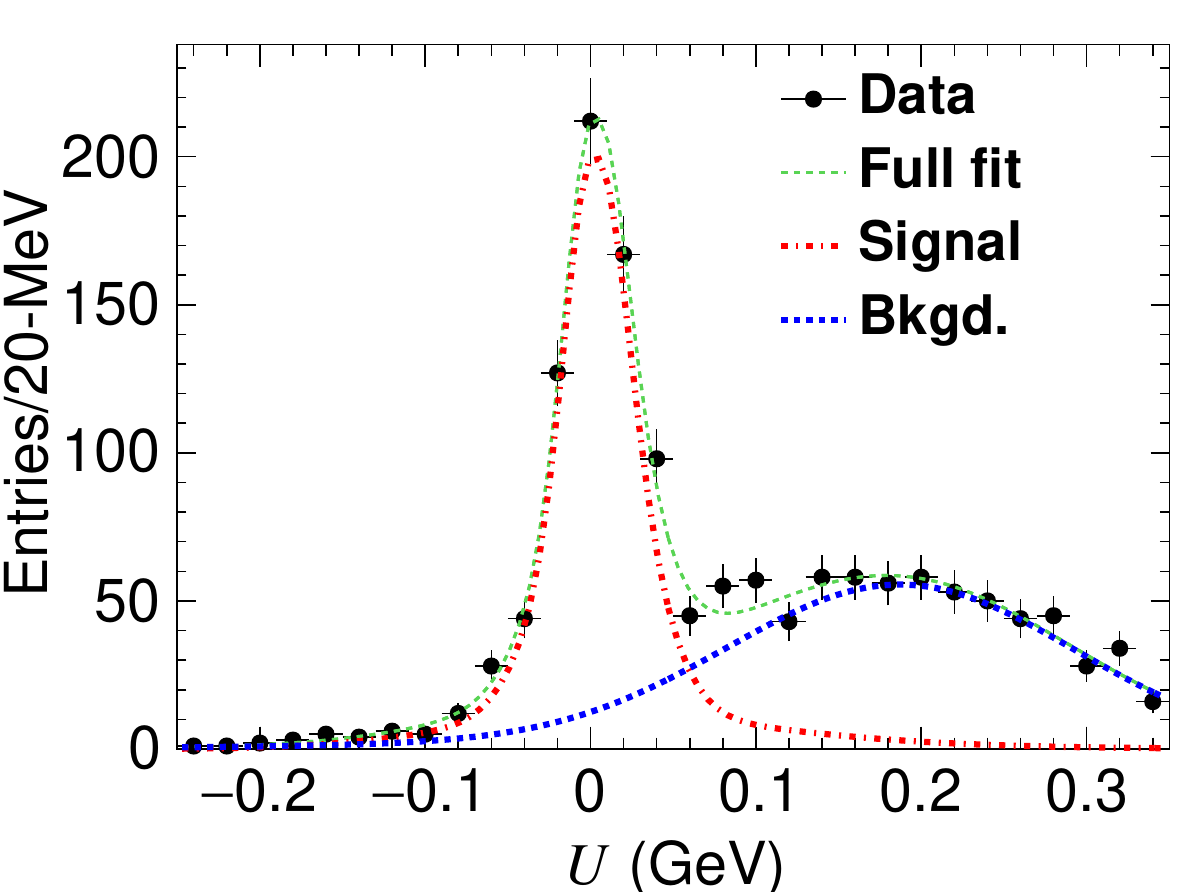}
\caption[]{Fits for mode~0 in Table~\ref{table:modes}, integrated over $\qsq$ and $\ctl$: (left) shows the fit to the signal component in {\tt GENBB} simulation using the template in Eq.~\ref{eqn:bGaus_form}; (middle) shows the fit to the background component in {\tt GENBB} simulation using Eq.~\ref{eqn:bkgd_form}; (right) shows the fit to the data. The individual two-piece Gaussian (TP Gauss) components are also shown.}
\label{fig:sigbkgd_swt_tweak90_tweaksig90_mode0}
\end{figure*}

Several techniques have been presented in the literature to perform background subtraction, the most common ones being sideband subtraction~\cite{Aubert:2004cp} and sWeighting~\cite{Pivk:2004ty,Xie:2009rka}. For amplitude analyses with a relatively large background, the effect of the sideband subtraction procedure on the derived uncertainties in the fit parameters was highlighted in Appendix~A of Ref.~\cite{Aubert:2004cp} and Sec.~XI.C of Ref.~\cite{spd-paper}. The sWeighting method leads to similar problems with the fit parameter uncertainties, in addition to the fact that the sWeights can be negative. Therefore, ad hoc scale factors are sometimes added to the minimization function to scale the statistical uncertainties, for example as in Ref.~\cite{Aaij:2019mhf}. In this analysis, a novel background separation technique is adopted that leads to positive signal weights and retains all multi-dimensional correlations among the event variables.

Figure~\ref{fig:dlnu_stacked} shows the breakdown of the $\bdlnu$ data composition after all selection requirements, integrated over all the ten reconstruction modes in Table~\ref{table:modes}. The black filled circles are the data, while the stacked histograms are based on the {\tt GENBB} simulation sample, weighted to match the data luminosity. No fits in the discriminating variable $U$ have been performed at this stage. The main purpose of Fig.~\ref{fig:dlnu_stacked} is to identify the background sources. The primary source of background for this analysis is feed-down from $\overline{B} \rightarrow D^\ast \ell^- \overline{\nu}_\ell$, with the subsequent decay $D^\ast \to D \pi$ or $D^\ast \to D \gamma$ in the case of the neutral $D^0$. The $D^\ast$'s, being vector mesons, have a characteristic forward-angle peak~\cite{Richman:1995wm} as $\ctl \to 1$. The remaining small background in Fig.~\ref{fig:dlnu_stacked} mostly comprises charmless hadronic $B$ decay components as well as some contribution from $q\overline{q}$ continuum. In general, both the shape and scale of the backgrounds are dependent on the phase space variables $\phi \in \{\qsq, \ctl\}$ and the reconstruction mode. It is, therefore, necessary to perform signal-background separation in small $\Delta \phi \equiv \{\Delta(\qsq), \Delta(\ctl) \}$ bins, independently for each of the ten reconstruction modes.

\subsection{Setup and sample global fits}
\label{sec:swt_fits}

The signal and background lineshapes in the $U$ variable distributions are derived from the {\tt GENBB} simulation samples employing the truth-matched and non-truth-matched components, respectively. The lineshapes are constructed from a two-piece Gaussian template, defined as 
\begin{linenomath}
\begin{align}
\label{eqn:bGaus_form}
f_i(x; \mu_i, \sigma_{L,i}, \sigma_{R,i}, N_i) = \phantom{spacespacespacespace}  \nonumber \\
N_i \times \begin{cases} e^{- (x - \mu_i)^2 /2\sigma^2_{L,i}}, & \mbox{for } x \leq \mu_i  \\
e^{- (x - \mu_i)^2 /2\sigma^2_{R,i}}, & \mbox{for } x > \mu_i.
\end{cases}  
\end{align}
\end{linenomath}
The signal lineshape is a sum of four two-piece Gaussian functions, two central peaks ($i=0,1$) and two tails ($i=2,3$) on each side of $U=0$:
\begin{linenomath}
\begin{align}
\label{eqn:signal_form}
\cal{S} &\equiv N_s \left (\sum_{i=0,1,2,3} \alpha_i e^{- (x - \mu_i)^2 /2\sigma^2_{L,R,i}} \right),
\end{align}
\end{linenomath}
where $\sigma^2_{L,R,i}$ represent the widths of the two-piece Gaussian functions defined in Eq.~\ref{eqn:bGaus_form}. The $\alpha_i$'s are the relative fractions with $\alpha_0=1$ for the first central Gaussian. The overall pre-factor $N_s$ is left unconstrained in all fits. 

Similarly, the background lineshape, $\cal{B}$, is templated using two two-piece Gaussian functions with $\mu_i$ shifted away from $U=0$, so that the signal and background lineshapes have well-demarcated and disjoint shapes:
\begin{linenomath}
\begin{align}
\label{eqn:bkgd_form}
\cal{B} &\equiv N_b \left (\sum_{j=0,1} \alpha_j e^{- (x - \mu_j)^2 /2\sigma^2_{L,R,j}} \right),
\end{align}
\end{linenomath}
with $\alpha_0=1$.  

For fits to the data, the normalizations of the signal and background components are always left unconstrained. For the signal component in Eq.~\ref{eqn:signal_form}, the shapes of the tails, $\{\mu_i, \sigma_{L,R,i}\}$ for $i\in\{2,3\}$, are kept fixed to the values obtained from fits to truth-matched signal in the {\tt GENBB} simulation, since these parts of the signal lineshape are away from the central peak and they cannot reliably be estimated from the data. For the rest of the nine parameters, $\{\alpha_{1,2,3}, \mu_{0,1}, \sigma_{L,R,0,1}\}$, the values in the data fits are constrained between $[(1-\kappa), 1/(1-\kappa)]$ times the nominal value obtained from the truth-matched simulation fit. For the background templates, all seven shape parameters in Eq.~\ref{eqn:bkgd_form} are allowed to vary between $[(1-\kappa), 1/(1-\kappa)]$ times the nominal values obtained from the non-truth-matched simulation (background) fit. Different choices of $\kappa\in [0,5\%,30\%]$ are studied to account for possible differences in the lineshapes between data and simulation, described further in Sec.~\ref{sec:syst_uncer}.

Figure~\ref{fig:sigbkgd_swt_tweak90_tweaksig90_mode0} shows sample fits for mode~0 in Table~\ref{table:modes} integrated over $\qsq$ and $\ctl$. The left panel shows the fit to the simulated signal, while the middle panel shows the fit to simulated background events. The lineshapes follow the templates in Eqs.~\ref{eqn:bGaus_form} and \ref{eqn:bkgd_form}. The individual two-piece Gaussian components are also shown. The right panel shows the fit to the data, validating the general procedure. Similar global fit quality checks were performed for the rest of the ten signal modes.

\subsection{Fits in local phase space regions}
\label{sec:local_phsp_binned_fits}

\begin{figure}
\centering
\includegraphics[width=1.68in]{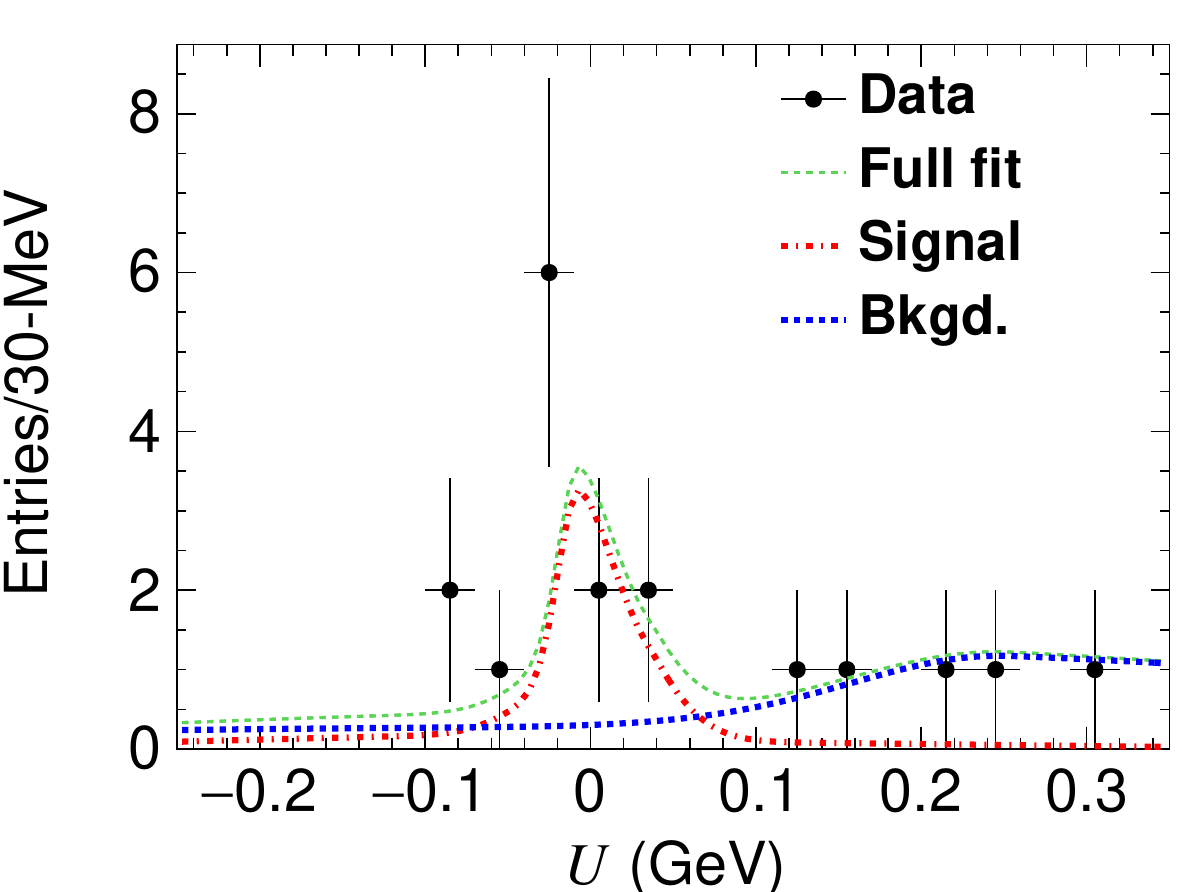}
\includegraphics[width=1.68in]{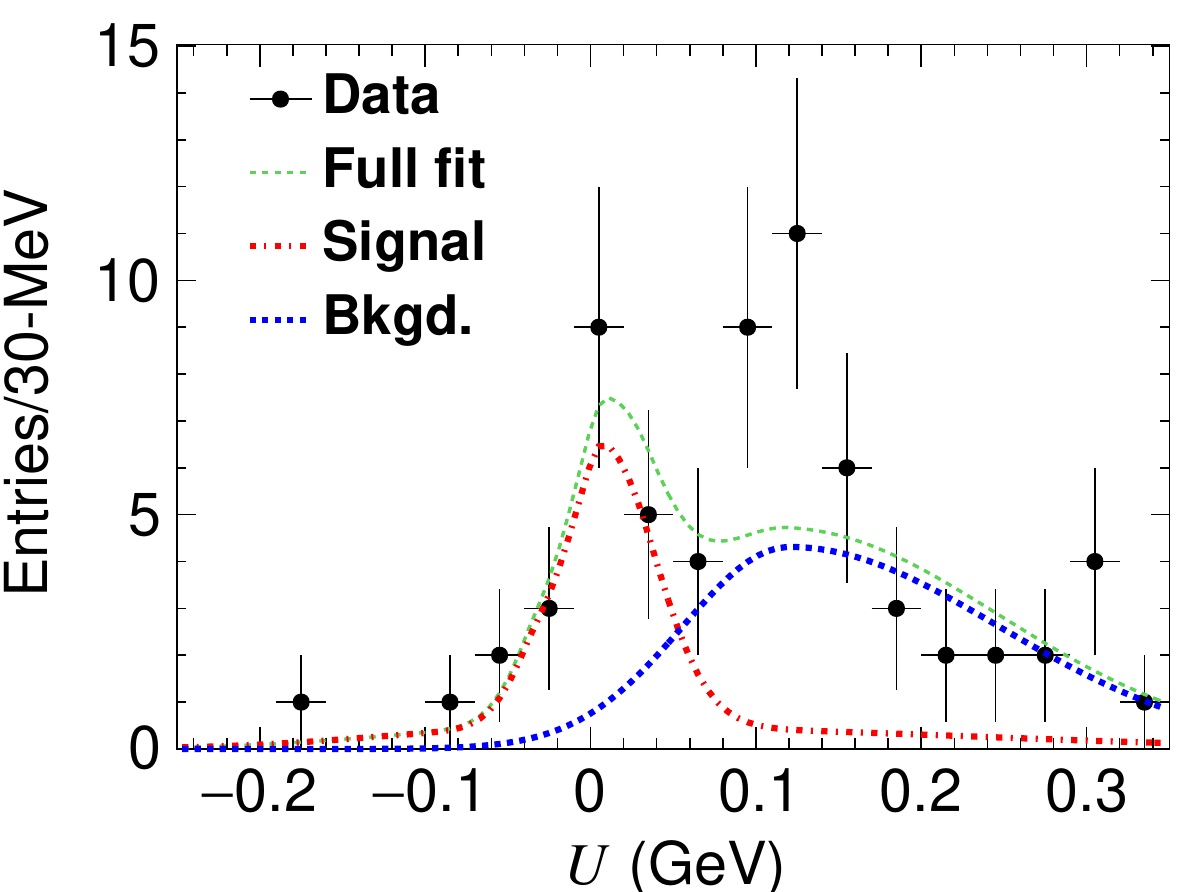}
\caption[]{Fits to data in two $\ctl$ bins for mode~3: (left) $|\ctl+0.85|<0.05$ and (right) $|\ctl-0.85|<0.05$. The lineshapes are taken from fits to simulation in the corresponding $\ctl$ bin. The background lineshapes vary strongly between the two $\ctl$ bins.}
\label{fig:sigbkgd_ctlbins}
\end{figure}

\begin{figure}
\centering
\includegraphics[width=1.68in]{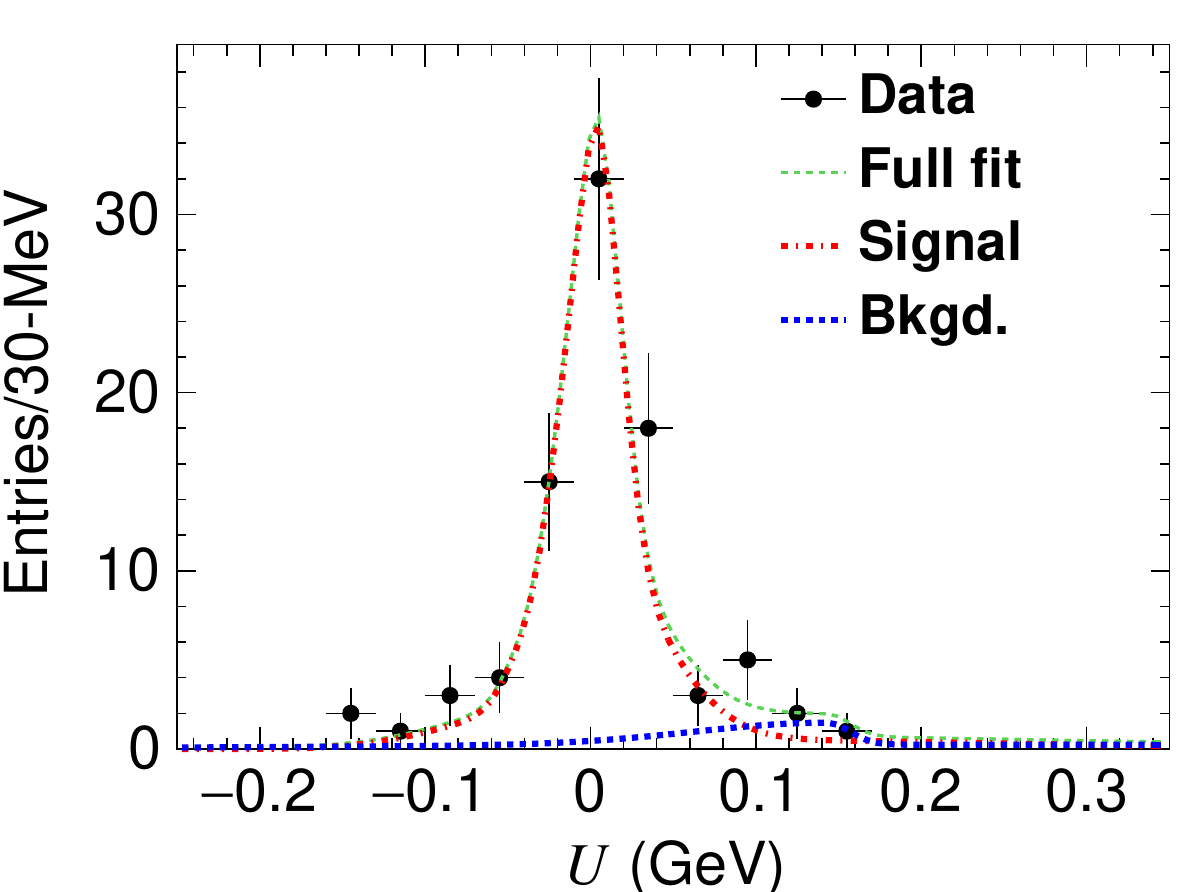}
\includegraphics[width=1.68in]{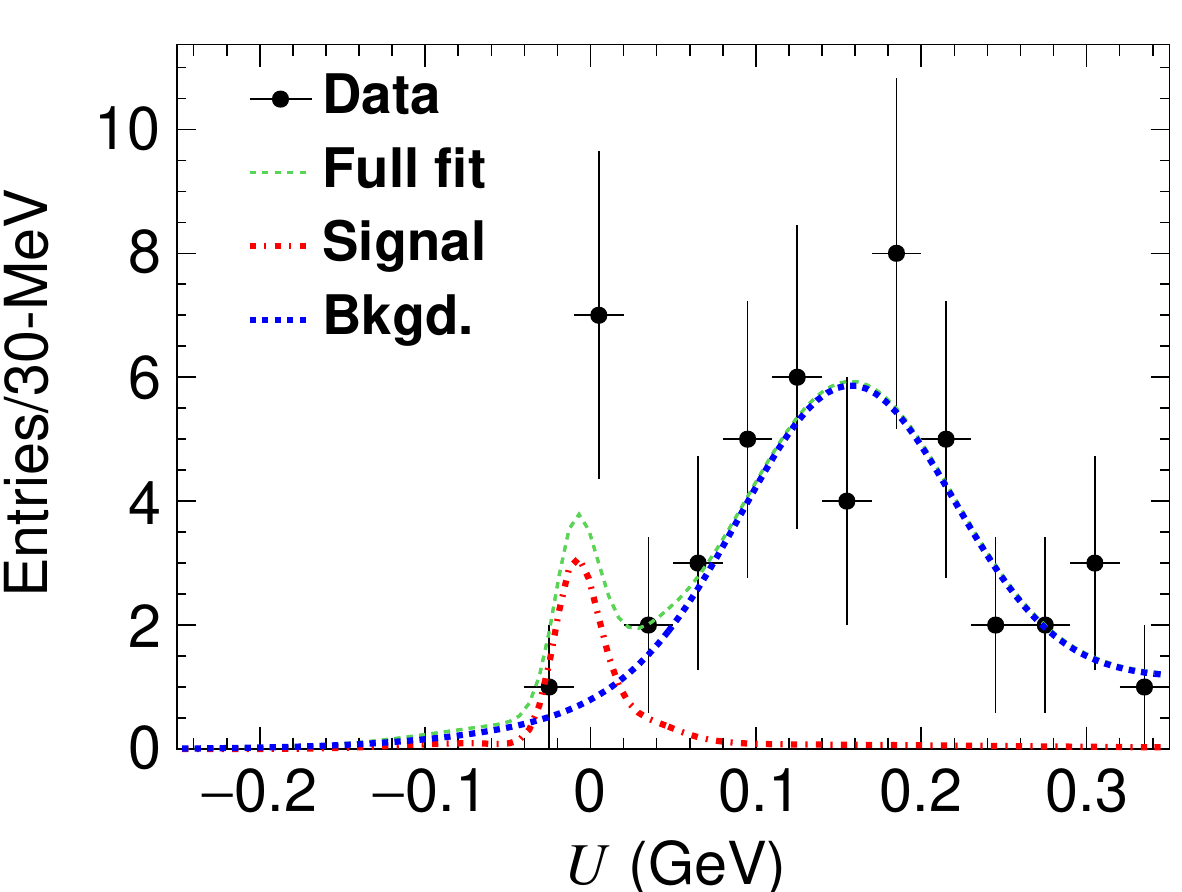}
\caption[]{Fits to data in two $\qsq$ bins for mode~2: (left) $|\qsq-0.75|<0.25$~GeV$^2$ and (right) $|\qsq-9.75|<0.25$~GeV$^2$. The lineshapes are taken from fits to simulation in the corresponding $\qsq$ bin, and vary strongly between the two $\qsq$ bins.}
\label{fig:sigbkgd_q2bins}
\end{figure}

For a two-dimensional angular analysis over the entire $\{\qsq, \ctl\}$ phase space, a single global background-separation fit, such as employed in the sWeighting~\cite{Pivk:2004ty} method, encounters difficulties. The signal and background lineshapes vary in phase space, particularly when close to the phase space boundaries, $\qsq\to q^2_{\rm \small max,min}$ and $\ctl \to \pm 1$. Figure~\ref{fig:sigbkgd_ctlbins} shows the fits in two $\ctl$ regions for mode~3, with the signal and background lineshapes derived from fits to the simulation, as discussed in Sec.~\ref{sec:swt_fits}. The background shape varies across the phase space, as can be expected from the fact that the physics backgrounds (such as $D^\ast$ feed-down) are phase-space dependent. Similarly, Fig.~\ref{fig:sigbkgd_q2bins} shows the fits in two $\qsq$ regions for mode~2. Close to $\qsq\approx 0$ there is also a kinematic supression in the high $U$ region that shapes the templates, since at low $\qsq$, the di-lepton breakup momentum is small which constrains the kinematically allowed range of $U$. 

The above features are demonstrated in Figs.~\ref{fig:sigbkgd_ctlbins} and \ref{fig:sigbkgd_q2bins} for pathological phase-space boundary regions in two modes. Similar features appear for all the ten modes. Detailed checks, as described above for the fits shown in Fig.~\ref{fig:sigbkgd_swt_tweak90_tweaksig90_mode0} are repeated in small phase space regions for each of the ten modes.  The checks demonstrate that within the statistical precision of the data, the signal and background lineshapes from simulation have the flexibility to provide good descriptions of the data

\subsection{Execution of the procedure in a continuous fashion}
\label{sec:qval}

The above method of performing fits in local phase space regions can be extended from a binned to a continuous procedure. For the $i^{th}$ event, an $N_c$ number of close-neighbor events in phase space are considered. To refine the notion of ``closeness'', the following ad hoc distance metric is defined between the $i^{th}$ and $j^{th}$ events in phase space:
\begin{linenomath}
\begin{align}
\label{eqn:metric}
g^2_{ij} = \displaystyle \sum_{k=1}^n \left[ \frac{\phi^i_k - \phi^j_k}{r_k}\right]^2,
\end{align}
\end{linenomath}
where $\vec{\phi}$ represents the $n$ independent kinematic variables in phase space, and $\vec{r}$ describes the corresponding ranges for normalization ($r_{\qsq}=10$~GeV$^2$, and $r_{\ctl}=2$ and $n=2$). The $N_c +1$ events are then fitted to a signal $\mathcal{S}(U)$ plus a background function $\mathcal{B}(U)$, of the same form as in Sec.~\ref{sec:swt_fits}. Once the functions $\mathcal{S}_i(x)$ and $\mathcal{B}_i(x)$ have been obtained from this fit for the $i^{th}$ event, the event is assigned a signal quality factor $Q_i$ given by:
\begin{equation}
Q_i = \frac{\mathcal{S}_i(U_i)}{\mathcal{S}_i(U_i) + \mathcal{B}_i(U_i)}. 
\label{eqn:qval_defn}
\end{equation}
The $Q$-factor is then used to weight the event's contribution for all subsequent calculations. For example, the total signal yield is simply defined as
\begin{linenomath}
\begin{equation} 
\mathcal{Y} = \sum_i Q_i.
\end{equation} 
\end{linenomath}

This method has already been applied to multi-dimensional angular analyses elsewhere with excellent results~\cite{Dey:2014tfa, Williams:2009ab, Williams:2009yj, Dey:2010hh}. In the context of heavy quark physics, similar background subtraction schemes in small phase space bins around the given event have also been studied in the context of semileptonic $D$ decay processes at FOCUS, CLEO, and BESIII~\cite{cleo_fit_method,Liu:2013zx, CLEO:2011ab}. Each of the five $D$ meson decay modes listed in Table~\ref{table:modes} as well as the two different lepton samples ($e/\mu$) are processed separately since each of the ten resulting categories have different signal-background characteristics. The fit framework remains the same as in Sec.~\ref{sec:swt_fits}. 

\begin{figure*}
\centering
\includegraphics[width=6.5in]{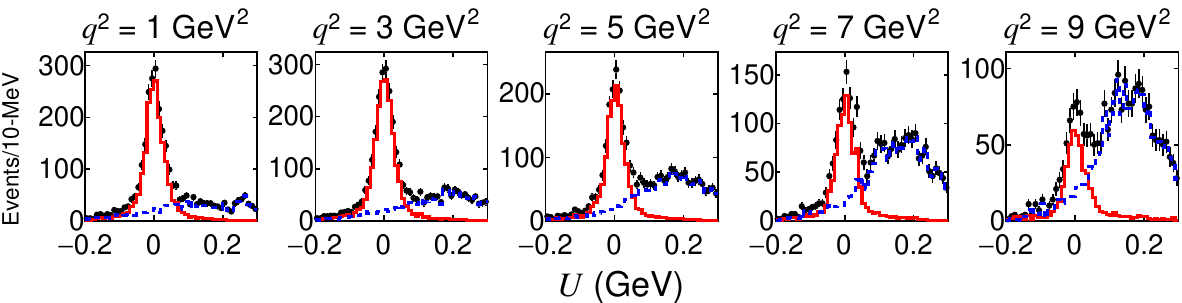}
\includegraphics[width=6.5in]{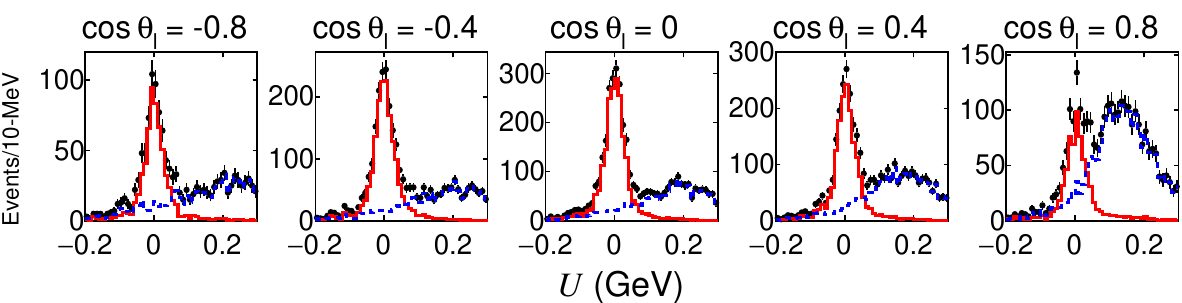}
\caption[]{Results from the local fits in (top) $\qsq$ and (bottom) $\ctl$ bins. The red and blue points are the yields obtained using $Q_i$ and ${(1-Q_i)}$ as weights for the $i^{th}$ event, representing the signal and background components, respectively. The $\qsq$ and $\ctl$ bin centers are marked on the panels. The bin widths are 2~GeV$^2$ in $\qsq$ and 0.2 units in $\ctl$. The plots combine results from all the ten modes.}
\label{fig:sigbkgd_dlnu}
\end{figure*}

Making a judicious choice for $N_c$ is based on two opposing constraints -- a high value of $N_c$ integrates over a large phase space region, while a too small value results in too few events to perform a fit. The total number of events, including signal and background for all the 10 modes is 16\,701. The nominal choice of $N_c=50$ is found to give stable fits for all events and amounts to around $\sqrt{16\,701 / 50 / 10}\sim 6$ effective ``bins'' in each of the $\phi$ dimensions.

Figure~\ref{fig:sigbkgd_dlnu} shows the results using $N_c = 50$, integrated over all the ten modes, and the signal and background shapes fixed to the simulation in the individual event-wise fits. In each panel, the black points show the total yields. The red and blue points represent the signal and background components, respectively. There are several noteworthy facets of this signal extraction technique. Each event is processed independently. That is, the functions $\mathcal{S}(x)$ and $\mathcal{B}(x)$ in Eq.~\ref{eqn:qval_defn} are obtained event-by-event. Since these fits are performed in local phase space regions independently for each reconstruction mode, variations in the signal resolution and the background compositions are accounted for. The background levels increase as $\ctl \to +1$. This is because the $D^\ast$ feed-down is prominently forward-peaked in $\theta_\ell$. Similarly, the $k^3$ dependence in Eq.~\ref{eqn:rate_formula} strongly suppresses the rate for pseudoscalar mesons at larger $\qsq$, while for vector mesons, the rate is slightly peaked towards larger $\qsq$~\cite{Richman:1995wm}. Therefore, the $D^\ast$ feed-down increases with increasing $\qsq$ (with a fall off at the phase space edges). It is important to note that although the $\thetal$ and $\qsq$ parts factorize for the signal in Eq.~\ref{fig:sigbkgd_dlnu} (neglecting acceptance effects), they are strongly correlated in the background. Therefore, even though one-dimensional projections are shown in Fig.~\ref{fig:sigbkgd_dlnu}, the signal-background separation is a two-dimensional problem. 

\subsection{Final yields after \boldmath${|U|\leq 50}$~MeV requirement}
\label{sec:final_yields}

After the $Q$-factors have been extracted for each event, a final $|U|\leq 50$~MeV selection requirement is placed to truncate the sidebands where the signal-background separation is less reliable. This not only ensures that selected events are around the $|U| \to 0$ region, corresponding to well-reconstructed events, but also avoids systematic uncertainties arising from modeling the long tail at large $U$ due to undetected soft photons. Henceforth, the following additional selection criteria are applied: $|\ctl| < 0.97$ and $\qsq \in [0.5,10]$~GeV$^2$, thereby trimming the phase space edges. For $B\to D$, $q^2_{\rm max} \sim 11.6$~GeV$^2$. However, from the $k^3$ dependence in Eq.~\ref{eqn:rate_formula}, the rate decreases rapidly as $\qsq \to q^2_{\rm max}$, so that there are only very few signal events in this region. Additionally, lattice QCD results are most precise here as well, so that the data do not add much information, comparatively. This further motivates limiting the upper $\qsq$ range to 10~GeV$^2$. Table~\ref{table:modes} lists the final yields for each $D$ meson decay mode after all selection requirements and signal-background separation. About 5500 signal events are available for the final amplitude analysis.

\section{Unbinned angular fits}

\subsection{The negative log-likelihood with acceptance correction and background subtraction}
\label{sec:uml_fits}

\subsubsection{\babar-only ``non-extended'' contribution}

Following the formalism described in Ref.~\cite{chung1993formulas}, the probability density function (pdf) for detecting an event within the phase-space element $[\phi, \phi +\Delta\phi]$ is
\begin{linenomath}
\begin{equation}
\mathcal{P}(\vec{x}, \phi) = \displaystyle \frac{ \displaystyle \frac{\deriv N(\vec{x}, \phi)}{\deriv \phi} \eta(\phi) \Delta \phi}{ \displaystyle \int \frac{\deriv N(\vec{x}, \phi)}{\deriv\phi} \eta(\phi) \deriv\phi},
\end{equation}
\end{linenomath}
where $\deriv N(\vec{x}, \phi)/\deriv\phi$ is the rate term, $\eta(\phi)$ is the phase-space dependent detector efficiency or acceptance, and $\vec{x}$ denotes the relevant set of fit parameters that the differential rate depends on. The normalization integral constraint (for pure signal)~\footnote{In the following notation, $\deriv N/\deriv \phi \equiv \deriv N(\vec{x}, \phi)/\deriv \phi$ is implied.}
\begin{linenomath}
\begin{equation}
\label{eqn:norm_int_def}
\mathcal{N}(\vec{x}) = \int \frac{\deriv N}{\deriv\phi} \eta(\phi) \deriv \phi \equiv \bar{N}(\vec{x}) = N_{\text{\scriptsize data}} 
\end{equation}
\end{linenomath}
ensures that the pdf is properly normalized to unity. The estimated yield (from the fit), $\bar{N}(\vec{x})$, is equal to the actual measured yield \footnote{Strictly speaking, this should be equal to the average experimental yield upon repeating the experiment many times.}. The ``non-extended'' likelihood function is then defined as
\begin{linenomath}
\begin{equation}
 \label{eqn:likelihood_def}
 \mathcal{L}(\vec{x}) = \prod\limits_{i=1}^{N_{\text{\scriptsize data}}} \mathcal{P}(\vec{x},\phi_i).
\end{equation}
\end{linenomath}
The likelihood function is insensitive to the overall scale of the rate function, since this cancels in the pdf definition. The objective of the angular fit is to maximize the likelihood as a function of the fit parameters $\vec{x}$, equivalent to minimizing the negative log likelihood (NLL). For the likelihood function in Eq.~\ref{eqn:likelihood_def}, the NLL reads
\begin{linenomath}
\begin{align}
\label{eqn:log_likelihood}
  -\ln{\mathcal{L}}(\vec{x}) =& - \sum\limits_{i=1}^{N_{\text{\scriptsize data}}} \ln{\mathcal{P}(\vec{x},\phi_i)} \nonumber \\
  \simeq& \displaystyle N_{\text{\scriptsize data}}\,\ln \left[\mathcal{N}(\vec{x})\right] - \sum\limits_{i=1}^{N_{\text{\scriptsize data}}} \ln \left[ \frac{\deriv N}{\deriv\phi} \eta(\phi) \right]_i.
\end{align}
\end{linenomath}

In Eq.~\ref{eqn:log_likelihood}, as noted earlier, $\eta$ denotes the detector acceptance that depends on $\phi$. The acceptance is incorporated in the fit using the {\tt GENBB} simulation. The acceptance $\eta(\phi)$ is not known as an analytic function but enters into the normalization integral in Eq.~\ref{eqn:norm_int_def}. Using the approximation
\begin{linenomath}
\begin{equation}
  \mathcal{N} = \int \frac{\deriv N}{\deriv \phi} \eta(\phi) \deriv \phi \equiv \left(\int \deriv \phi\right) \left\langle  \frac{\deriv N}{\deriv \phi} \eta(\phi) \right\rangle,
\end{equation}
\end{linenomath}
the average efficiency-incorporated rate term can be calculated using $N_{\rm {\scriptsize sim}}^{\rm {\scriptsize gen}}$ simulation events (see Sec.~\ref{sec:sim_samples}) that are generated uniformly in $\phi$, as
\begin{linenomath}
\begin{equation}
   \left\langle  \frac{\deriv N}{\deriv\phi} \eta(\phi) \right\rangle = \sum_{i=1}^{N_{\text{\scriptsize sim}}^{\text{\scriptsize gen}}} \frac{\deriv N}{\deriv\phi} \frac{\eta(\phi)}{N_{\text{\scriptsize sim}}^{\text{\scriptsize gen}}} = \sum_{i=1}^{N_{\text{\scriptsize sim}}^{\text{\scriptsize acc}}} \frac{\deriv N}{\deriv\phi} \frac{1}{N_{\text{\scriptsize sim}}^{\text{\scriptsize gen}}},
\label{eqn:flat_MC_avg_rate}
\end{equation}
\end{linenomath}
where, in the last step, the acceptance is incorporated by summing only over the ``accepted'' simulation events after reconstruction and detector inefficiencies. That is, $\eta$ is either 1 or 0, the event being either reconstructed or not.

Ignoring terms that are not variable in the fit, for pure signal,
\begin{linenomath}
\begin{align}
-\ln \mathcal{L}(\vec{x}) = N_{\text{\scriptsize data}} \times \ln \left[\displaystyle \sum_{i=1}^{N_{\text{\scriptsize sim}}^{\text{\scriptsize acc}}} \frac{\deriv N}{\deriv\phi} \right] - \displaystyle \sum_{i=1}^{N_{\text{\scriptsize data}}} \ln \left[\frac{\deriv N}{\deriv\phi} \right]_i.
\label{eqn:log_likelihood1}
\end{align}
\end{linenomath}
The background subtraction procedure is made explicit in Eq.~\ref{eqn:log_likelihood1} by weighting the data terms by their corresponding $Q$-values as
\begin{linenomath}
\begin{align}
\label{eqn:log_likelihood_qval}
-\ln \mathcal{L}(\vec{x}) =& \left[ \displaystyle \sum_{i=1}^{N_{\text{\scriptsize data}}} Q_i\right]  \times \ln \left[\displaystyle \sum_{i=1}^{N_{\text{\scriptsize sim}}^{\text{\scriptsize acc}}} \frac{\deriv N}{\deriv\phi} \right] \nonumber
 \\
& - \displaystyle \sum_{i=1}^{N_{\text{\scriptsize data}}} Q_i \ln \left[\frac{\deriv N}{\deriv\phi} \right]_i,
\end{align}
\end{linenomath}
where $N_{\text{\scriptsize data}}$ refers to the number of events after all selections. Equation~\ref{eqn:log_likelihood_qval} assumss that the simulation is generated uniformly in the kinematic variables such that the expected rate could be directly incorporated in the NLL by weighting each simulation event by the rate function, as shown in Eq.~\ref{eqn:flat_MC_avg_rate}. However, the existing {\tt GENBB} simulation samples employ a generator that uses a quark-model-based FF calculation (ISGW2~\cite{isgw2}) for $f_+(\qsq)$ and generates events according to Eq.~\ref{eqn:rate_formula}. To convert the existing {\tt GENBB} simulation samples to a uniform generator model, the contribution from each accepted {\tt GENBB} simulation event to the NLL is given an additional weight factor
\begin{linenomath}
\begin{equation}
\tilde{w} = 1/\left[\frac{\deriv N}{\deriv\phi}\right]_{\mbox{\scriptsize ISGW2}}.
\label{eqn:isgw2_wt}
\end{equation}
\end{linenomath}
Therefore, the final expression for the NLL is
\begin{linenomath}
\begin{align}
-\ln \mathcal{L}(\vec{x})\LARGE|_{\tiny \babar} =& \left[ \displaystyle \sum_{i=1}^{N_{\text{\scriptsize data}}} Q_i\right]  \times \ln \left[\displaystyle \sum_{i=1}^{N_{\text{\scriptsize sim}}^{\text{\scriptsize acc}}} \tilde{w}_i \left(\frac{\deriv N}{\deriv \phi}\right)_i \right] \nonumber
 \\
& - \displaystyle \sum_{i=1}^{N_{\text{\scriptsize data}}} Q_i \ln \left[\frac{\deriv N}{\deriv\phi} \right]_i.
\label{eqn:log_likelihood_qval_wt}
\end{align}
\end{linenomath}
This reweighting assumes that the rate predicted by the generator model is not zero. If there are no events in a phase-space bin, reweighting or redistribution of events cannot work. To take this into account, as mentioned in Sec.~\ref{sec:final_yields}, fits are performed within the region $|\ctl| < 0.97$ and $\qsq \in [0.5,10]$~GeV$^2$; that is, truncating the phase-space edges. The NLL in Eq.~\ref{eqn:log_likelihood_qval_wt} is calculated for each mode individually and summed over the ten modes.

\subsubsection{External constraints}

Two types of external constraints are imposed. The NLL in Eq.~\ref{eqn:log_likelihood_qval_wt} using the \babar\ data is of the non-extended type and cannot set the overall normalization. To set the normalization of the FF's, the $w\to 1$ region calculations from lattice QCD~\cite{Lattice:2015rga} are added as Gaussian constraints. In addition, to access $|\Vcb|$, the absolute $\qsq$-differential rate data from Belle~\cite{Glattauer:2015teq} are also incorporated as external Gaussian constraints. The total minimization quantity is
\begin{linenomath}
\begin{align}
\mathbb{L}_{\rm \small total}(\vec{x}) =& -2\ln \mathcal{L}(\vec{x})\LARGE|_{\tiny \babar} + \chi^2(\vec{x})\LARGE|_{\rm \small Belle} \nonumber \\
& \hspace{2.45cm} + \chi^2(\vec{x})\LARGE|_{\rm \small FNAL/MILC},
\label{eqn:nll_tot}
\end{align}
\end{linenomath}
where the first term corresponds to the unbinned \babar\ NLL, while the second and third terms correspond to the Gaussian constraints due to the external inputs. The Belle-16~\cite{Glattauer:2015teq} data set comprises 40 $\deriv \Gamma/\deriv w$ data points, while the FNAL/MILC QCD~\cite{Lattice:2015rga} data set comprises 6 $f_{0,+}(w)$ data points. The covariance matrices for these external data sets allow construction of the two partial $\chi^2$ components, $\chi^2(\vec{x})\LARGE|_{\rm \small Belle}$ and $\chi^2(\vec{x})\LARGE|_{\rm \small FNAL/MILC}$, for a given set of fit parameters. The values of the partial $\chi^2$ components from the external constraints are reported in the fit results; however no $p$-values to these individual data sets are quoted, since the fit minimizes the full NLL in Eq.~\ref{eqn:nll_tot}.

\begin{table*}
   \caption{\label{table:bgl_results_N2} The $N=2$ BGL results including statistical uncertainties only. A version of the fit excluding the Belle~\cite{Glattauer:2015teq} results is also provided for comparison.}
  \begin{center}
     \begin{tabular}{ccccccccc} \\ \hline \hline
      fit configuration& $a^{f_+}_0\times 10$ & $a^{f_+}_1$ & $a^{f_+}_2$ & $a^{f_0}_1$ & $a^{f_0}_2$ & $|\Vcb|\times 10^3$ & $\chi^2_{\rm {\scriptsize MILC}}$ & $\chi^2_{\rm {\scriptsize Belle}}$\\ \hline
        \babar-1, Belle & $0.126 \pm 0.001$ & $-0.096 \pm 0.003$ & $0.352 \pm 0.052$ & $-0.059 \pm 0.003$ & $0.155 \pm 0.049$ & $41.09 \pm 1.16$ & $1.15$ & $24.50$\\
        \babar-2, Belle & $0.126 \pm 0.001$ & $-0.096 \pm 0.003$ & $0.352 \pm 0.052$ & $-0.059 \pm 0.003$ & $0.155 \pm 0.049$ & $41.12 \pm 1.16$ & $1.17$ & $24.54$\\ 
        \babar-3, Belle & $0.126 \pm 0.001$ & $-0.096 \pm 0.003$ & $0.350 \pm 0.052$ & $-0.059 \pm 0.003$ & $0.153 \pm 0.049$ & $41.12 \pm 1.16$ & $1.18$ & $24.55$\\ 
        \babar-4, Belle & $0.126 \pm 0.001$ & $-0.096 \pm 0.003$ & $0.352 \pm 0.052$ & $-0.059 \pm 0.003$ & $0.156 \pm 0.049$ & $41.05 \pm 1.17$ & $1.14$ & $24.45$\\
        \babar-1 & $0.126 \pm 0.001 $ & $-0.097 \pm 0.003$ & $0.334 \pm 0.063$ & $-0.059 \pm 0.003$ & $0.133 \pm 0.062$ & - & $1.55$ & -\\ \hline \hline
    \end{tabular}
   \end{center}
\end{table*}

\begin{table}
   \caption{ \label{table:bgl_results_N3} The $N=3$ BGL results with \babar-1 and without systematic uncertainties.}
  \begin{center}
     \begin{tabular}{cr} \\ \hline \hline
      variable  & value\phantom{dum}\\ \hline
      $a^{f_+}_0\times 10$ & $0.126 \pm 0.001$ \\
      $a^{f_+}_1$ & $-0.098 \pm 0.004$ \\
      $a^{f_+}_2$ & $0.626 \pm 0.241$ \\
      $a^{f_+}_3$ & $-3.939 \pm 3.194$ \\
      $a^{f_0}_1$ & $-0.061 \pm 0.003$ \\
      $a^{f_0}_2$ & $0.435 \pm 0.205$ \\
      $a^{f_0}_3$ & $-3.977 \pm 2.840$ \\
      $|\Vcb|\times 10^3$ & $40.74 \pm 1.18$ \\
      $\chi^2_{\rm {\scriptsize FNAL/MILC}}$ & $0.001$\phantom{aaa} \\
      $\chi^2_{\rm {\scriptsize Belle}}$ & $23.68$ \phantom{aa}\\ \hline \hline
    \end{tabular}
   \end{center}
\end{table}

\subsubsection{Fit configurations}
\label{sec:fit_configs}

The nominal fit results are provided using $Q$-factors with $N_c=50$ and fixing the signal and background shapes $U$ (locally in phase-space and not globally) according to the simulation. For the lattice results, including the synthetic data from HPQCD~\cite{Na:2015kha} as provided in Ref.~\cite{Glattauer:2015teq} leads to covariance matrices not being positive definite, while the effect on the mean values of the fit results are negligible, since the HPQCD uncertainties are much larger than those from FNAL/MILC~\cite{Lattice:2015rga}. Hence, only the FNAL/MILC~\cite{Lattice:2015rga} lattice QCD calculations are used. 

For the CLN fits using Eq.~\ref{eq:cln}, only the $f_+$ part of the FNAL/MILC~\cite{Lattice:2015rga} calculations are employed. For the BGL fits, for the \babar\ data part, the appropriate masses of the $B$ and $D$ mesons are employed for each mode in the conversion between the $\qsq$, $w$, and $z$ variables in Eqs.~\ref{eqn:w_def} and~\ref{eq:bgl_z_expan}. For the FNAL/MILC data and in employment of the kinematic relation in Eq.~\ref{eqn:maxrecoilrel}, the masses are taken corresponding to the $B^-\to D^0\ellm\barnuell$ decay. The BGL expansion is truncated at $N=2$ both for $f_0$ and $f_+$ (three parameters each), so that there are five FF fit parameters while $a_0^{f_0}$ is derived from the five other parameters using Eq.~\ref{eqn:maxrecoilrel}. Cubic forms ($N=3$) of the BGL expansion are also investigated. However, with the present statistical precision, the highest order terms are found to have large uncertainties, leading to violation of unitarity conditions. Hence, only the $N=2$ BGL results are reported as the final results.

To consider systematic uncertainties, the \babar\ part of the fit includes four configurations for the background subtraction:
\begin{itemize}
\item \babar-1 (nominal), $N_c=50$, signal and background shapes locally fixed from simulation; 
\item \babar-2, $N_c=60$, signal and background shapes locally fixed from simulation; 
\item \babar-3, $N_c=50$, signal shapes allowed to vary by $5\%$ from the simulation;
\item \babar-4, $N_c=50$, tighter selection requirements ($\eex<0.6$~GeV, CL $>10^{-6}$).
\end{itemize}

\subsubsection{BGL results}
\label{sec:bgl_results}

Table~\ref{table:bgl_results_N2} reports the nominal $N=2$ BGL results including statistical uncertainties only, corresponding to the four background separation scenarios listed in Sec.~\ref{sec:fit_configs}. The $N=3$ results are reported in Table~\ref{table:bgl_results_N3}. The value of $\Delta(-2\ln \mathcal{L}(\vec{x})\LARGE|_{\tiny \babar})$ is found to be be zero between the $N=2$ and $N=3$ minimization points, signifying that the fit quality shows no improvement on addition of the cubic terms. In both cases, $\Delta(-2\ln \mathcal{L}(\vec{x})\LARGE|_{\tiny \babar})=1$ when the Belle component is included in the fit in Eq.~\ref{eqn:nll_tot}.

\begin{figure*}
\centering
\includegraphics[width=3.4in]{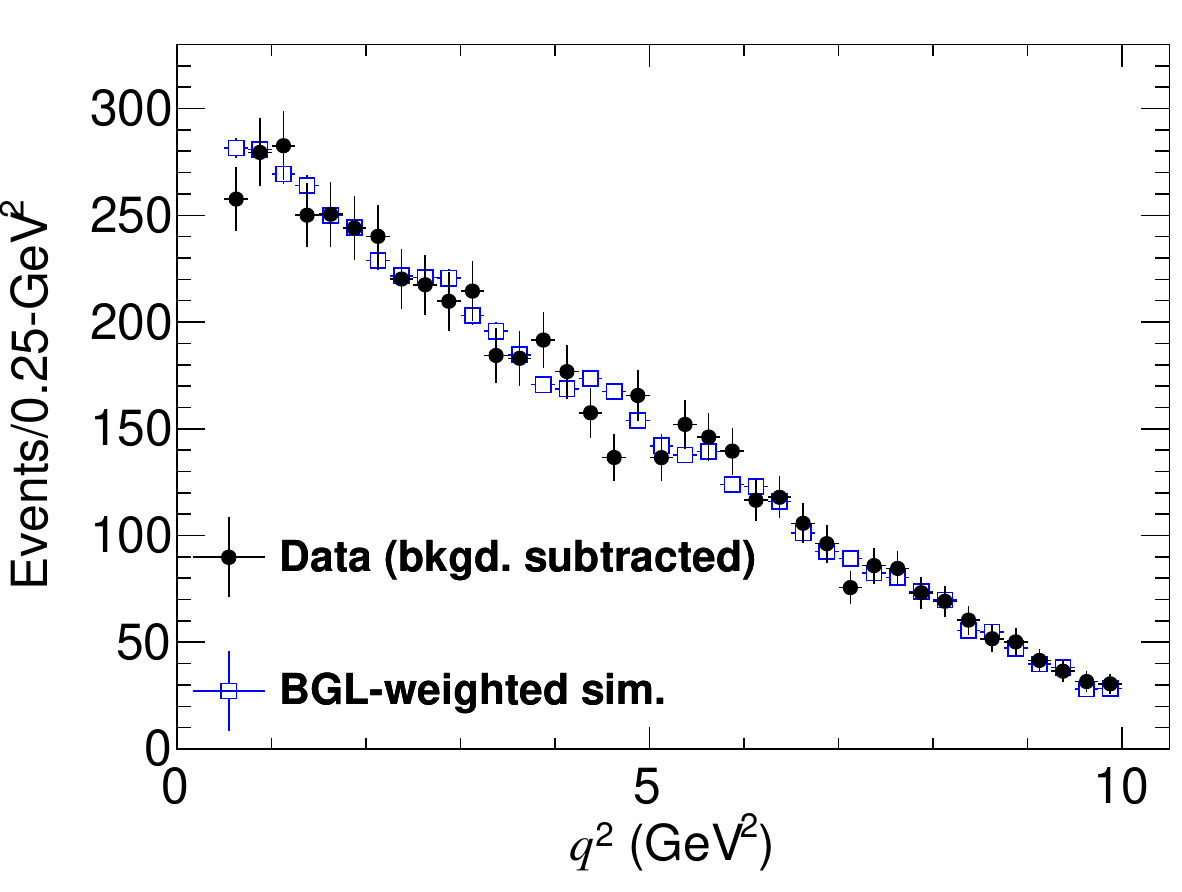}
\includegraphics[width=3.4in]{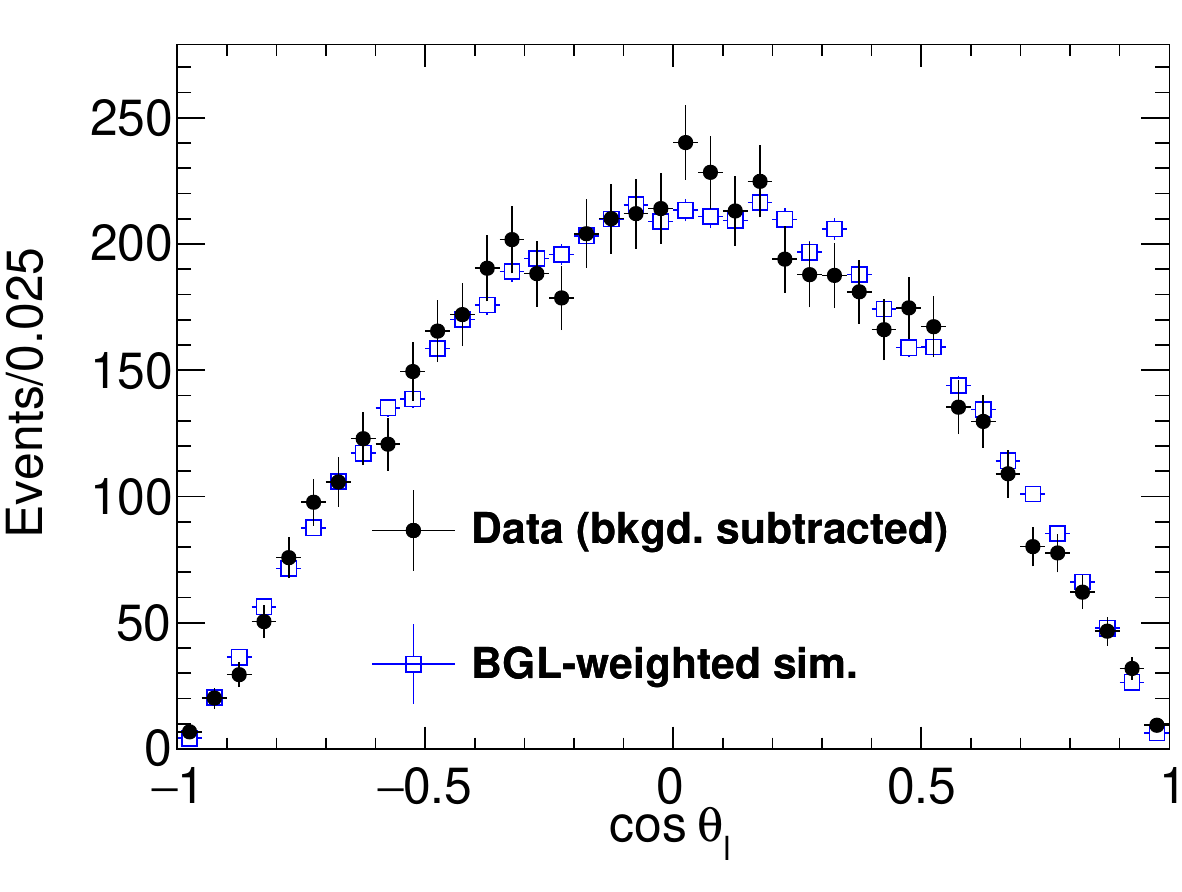}
\caption[]{Comparison between background-subtracted data (\babar-1 configuration) and the simulated events after acceptance effects and weighted by the BGL fit results as one-dimensional projections: (left) $\qsq$ and (right) $\ctl$.}
\label{fig:fit_quality_data_mc}
\end{figure*}

\subsubsection{CLN results}
\label{sec:cln_results}

\begin{table*}
  \begin{center}
   \caption{ \label{table:cln_results} The CLN results including statistical uncertainties only. A version of the fit excluding the Belle-16~\cite{Glattauer:2015teq} results is also provided for comparison.}
     \begin{tabular}{cccccc} \\ \hline \hline
      fit configuration & $\mathcal{G}(1)$ & $\rho^2_D$ &  $|\Vcb|\times 10^3$ & $\chi^2_{\rm {\scriptsize FNAL/MILC}}$ & $\chi^2_{\rm {\scriptsize Belle}}$\\ \hline
      \babar-1, Belle & $1.056 \pm 0.008$ & $1.155 \pm 0.023$ & $40.90 \pm 1.14$ & $1.04$ & $24.65$ \\ 
      \babar-2, Belle & $1.056 \pm 0.008$ & $1.156 \pm 0.023$ & $40.92 \pm 1.14$ & $0.99$ & $24.72$ \\ 
      \babar-3, Belle & $1.056 \pm 0.008$ & $1.156 \pm 0.023$ & $40.92 \pm 1.14$ & $1.00$ & $24.71$ \\
      \babar-4, Belle & $1.056 \pm 0.008$ & $1.154 \pm 0.023$ & $40.87 \pm 1.14$ & $1.09$ & $24.57$ \\ 
      \babar-1 & $1.053 \pm 0.008$ & $1.179 \pm 0.027$ & -- & $0.53$ & -- \\ \hline \hline
    \end{tabular}
   \end{center}
\end{table*}

Table~\ref{table:cln_results} lists the CLN results including statistical uncertainties only. The $\chi^2$ values against the binned FNAL/MILC~\cite{Lattice:2015rga} and Belle~\cite{Glattauer:2015teq} data are also reported. The FF slope $\rho^2_D$ tends to be slightly steeper than the current HFLAV (spring-21)~\cite{HFLAV21} average of $1.129 \pm 0.033$.

\subsubsection{Comparisons in $\qsq$ and $\ctl$}

Figure~\ref{fig:fit_quality_data_mc} shows the fit results as one-dimensional projections in $\qsq$ and $\ctl$, respectively. The black circles are the background-subtracted data, and the blue squares are the simulated events after acceptance, weighted by the BGL fit results. In particular, Fig.~\ref{fig:fit_quality_data_mc}b shows the $\ctl$ distribution, which exhibits the $\sin^2\thetal$ dependence expected in the SM. 

\section{Systematic uncertainties and final results}
\label{sec:syst_uncer}

\begin{figure}
\centering
\includegraphics[width=1.68in]{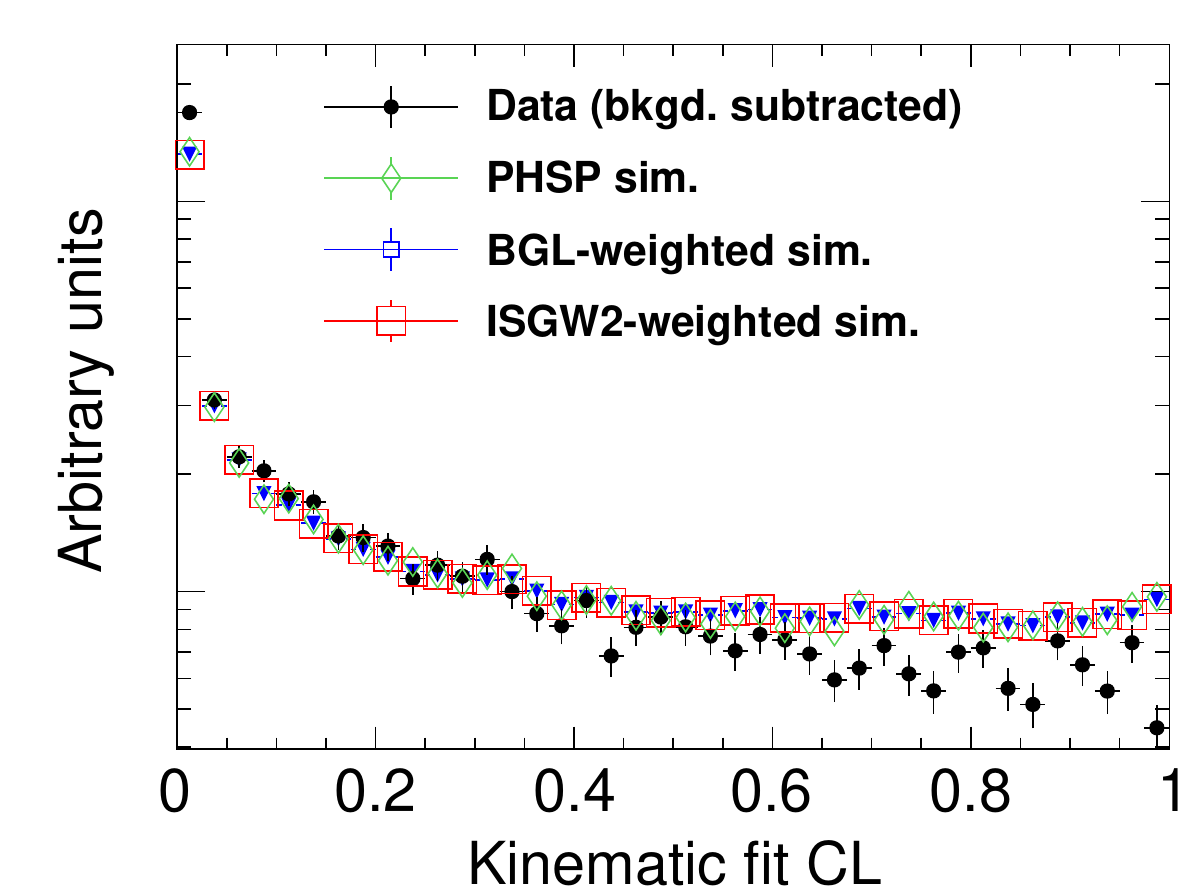}
\includegraphics[width=1.68in]{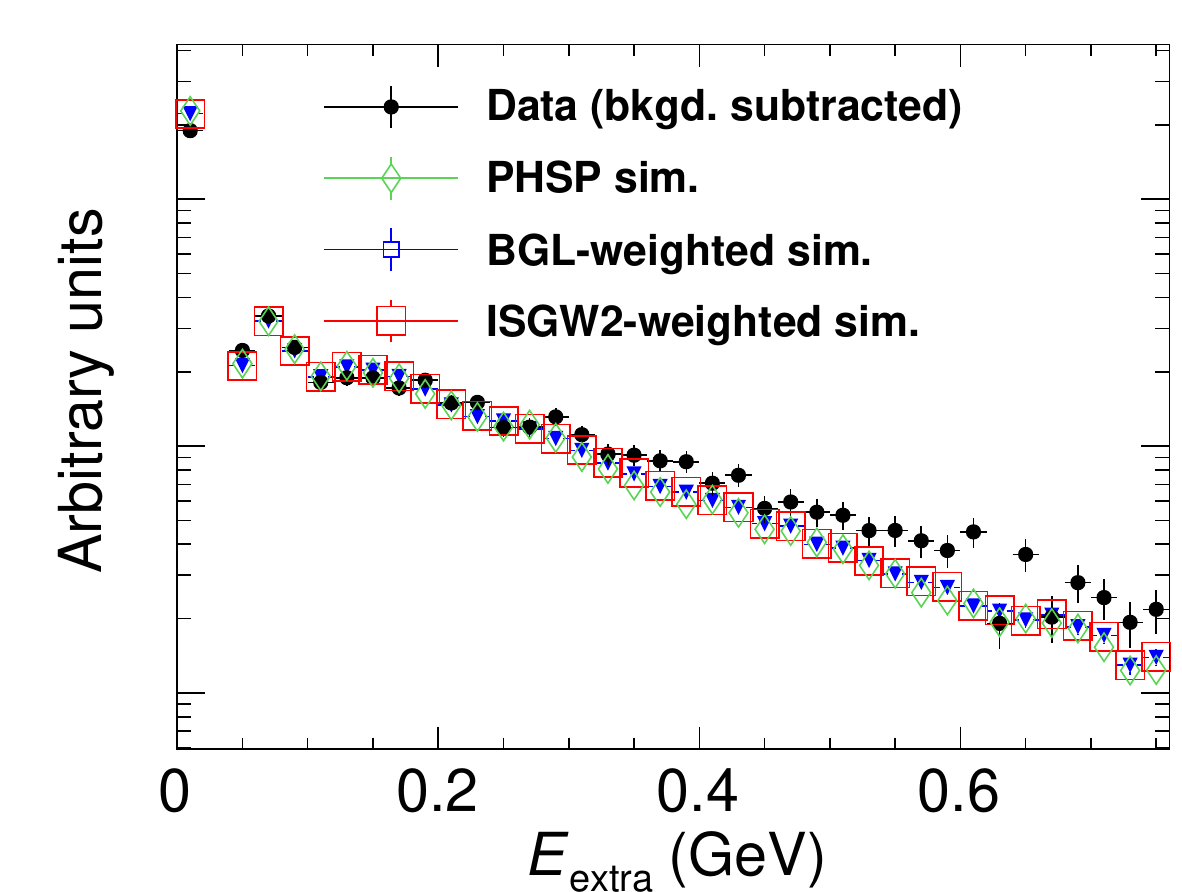}
\caption[]{Comparisons on a logarithmic scale between the background-subtracted data and the simulation in the selection variables: (left) CL from kinematic fit without the $U=0$ constraint and (right) $\eex$. The simulated events are either phase space (PHSP) or weighted according to the BGL and ISGW2 FF models.}
\label{fig:data_mc_prob_eex}
\end{figure}

Since the \babar\ part of the minimization function in Eq.~\ref{eqn:nll_tot} is of the non-extended type, uncertainties in knowledge of the \babar\ luminosity and individual $D$ meson decay mode branching fractions do not enter into the fit. Uncertainties in variables uncorrelated with the $\phi$ variables are also irrelevant for the angular analysis. The selection requirements in Sec.~\ref{sec:sel} are especially intended to be loose to reduce the possibilities of such correlations. Figure~\ref{fig:data_mc_prob_eex} shows the comparisons between background-subtracted data and the simulation. The mild differences seen are not correlated with the FF model, as verified by comparing distributions for the simulation using phase space (PHSP), the current BGL fit, and an older ISGW2~\cite{isgw2} FF models. Hence, no additional systematic uncertainty is assigned.

\begin{figure}
\centering
\includegraphics[width=1.65in]{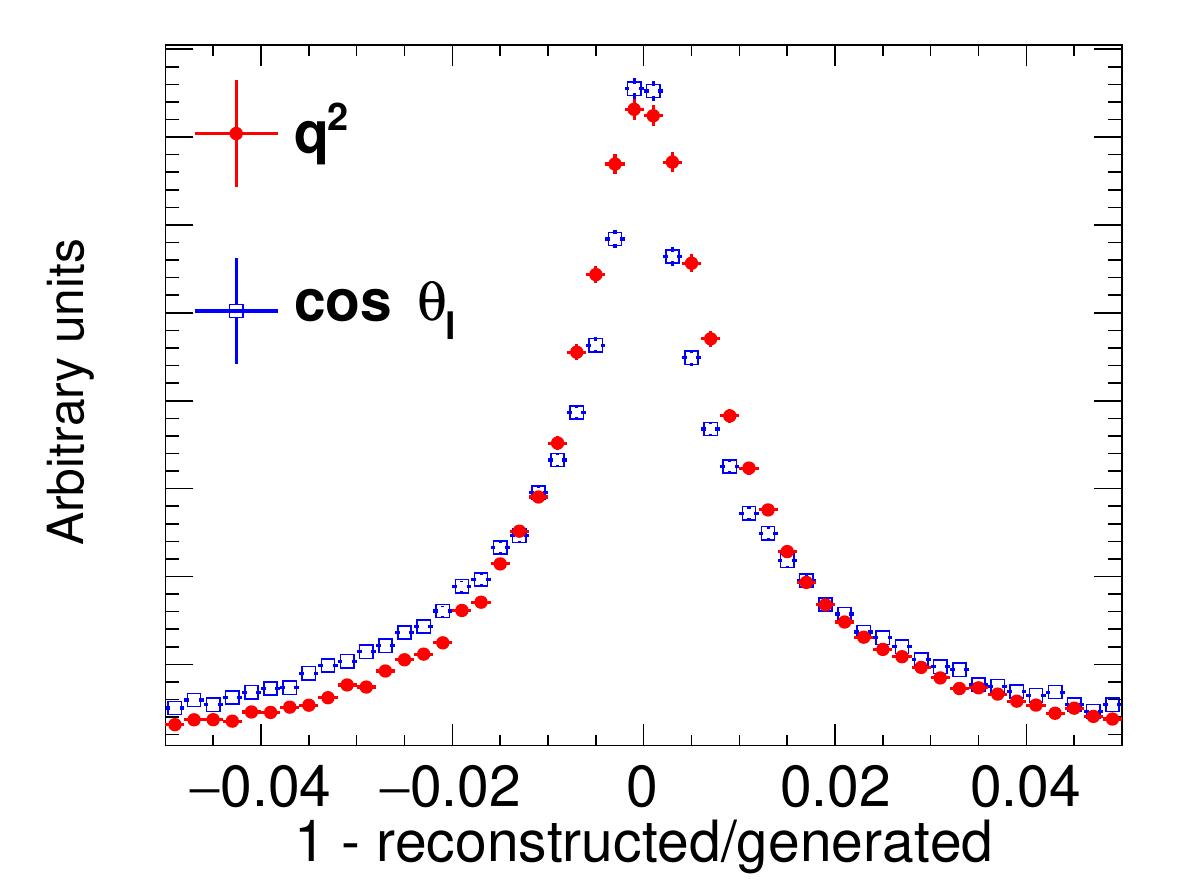}
\includegraphics[width=1.65in]{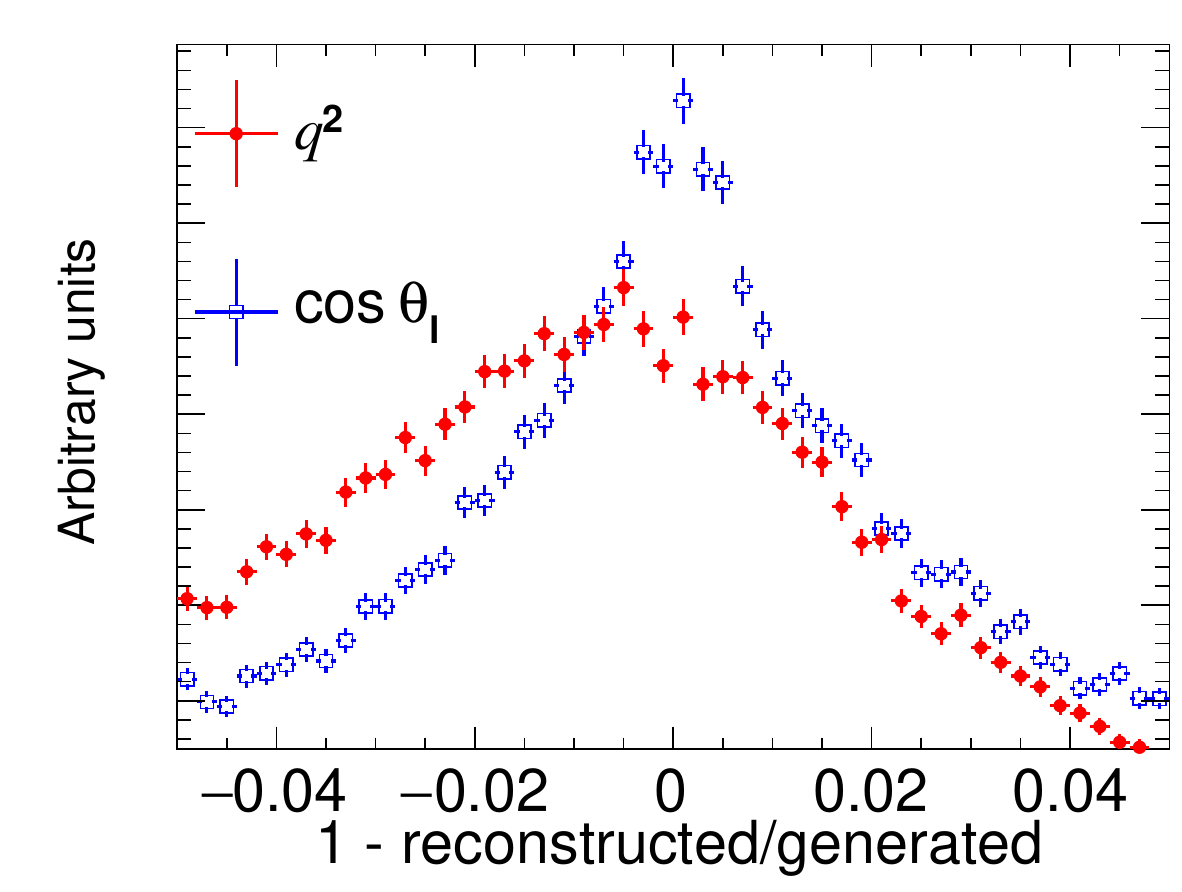}
\caption[]{The relative resolutions in the $\phi$ variables obtained from simulation: (left) using a kinematic fitting to zero missing mass and (right) no kinematic fit performed. The kinematic fit distinctly improves the resolutions from around $3.4\%$ to $2.6\%$.}
\label{fig:dlnu_q2_ctl_resolution}
\end{figure}

For correctly reconstructed variables, the ratio of the reconstructed-to-generated values should be close to unity. Figure~\ref{fig:dlnu_q2_ctl_resolution} shows the deviation of this ratio from unity, corresponding to the relative resolution in the $\phi$ variables. From the left panel in  Fig.~\ref{fig:dlnu_q2_ctl_resolution}, the highly constrained event topology and kinematic fitting result in excellent resolution, at the percent level. Adding the root-mean-squared distributions from each of the two histograms in the left panel of Fig.~\ref{fig:dlnu_q2_ctl_resolution}, the combined resolution in the kinematic variables is about $2.6\%$. The right panel in Fig.~\ref{fig:dlnu_q2_ctl_resolution} shows that this resolution degrades to about $3.4\%$ if the $\phi$ variables are constructed without a kinematic fit. The resolution effect is accounted for by evaluating the normalization integral in Eq.~\ref{eqn:flat_MC_avg_rate} with the reconstructed (instead of generated) kinematic variables. This procedure is appropriate up to second order effects from differences in the resolutions between data and simulation. To study the systematic uncertainty associated with the reconstruction, the fits are repeated employing the kinematic variables reconstructed without the kinematic fit. As a conservative estimate, the difference in results between these two fits is assigned as a systematic uncertainty.

\begin{figure}
\begin{center}
\includegraphics[width=3.4in]{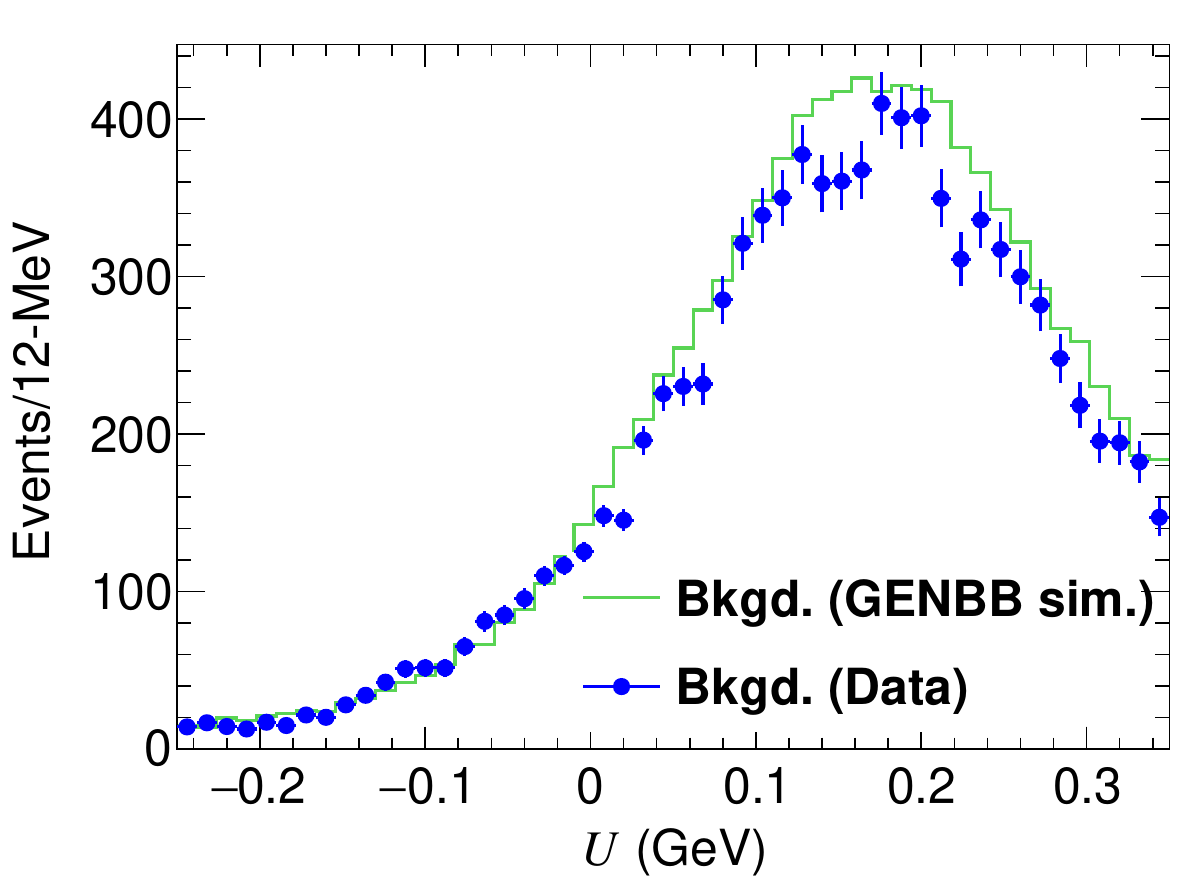}
\end{center}
\caption[]{Comparisons between the background distributions in the {\tt GENBB} simulation sample and background-subtracted data using $(1-Q)$ weights.}
\label{fig:dlnu_compare_genbb_data_bkgd}
\end{figure}

The other source of systematic uncertainty considered is the effect of the background subtraction. As described in Secs.~\ref{sec:cln_results} and~\ref{sec:bgl_results}, several variants of the background fits are employed. The maximum deviations of fit parameter values from the nominal outcomes are assigned as the systematic uncertainties. Figure~\ref{fig:dlnu_compare_genbb_data_bkgd} shows the comparisons between the background component in the {\tt GENBB} simulation sample and the background in the data, obtained from $(1-Q)$-weighted events. The mild differences away from the $U=0$ signal region indicate the imperfections in the {\tt GENBB} simulation, accounted for in the background-subtraction procedure, in a data-driven fashion.

To check for possible extremal differences in the background lineshapes between the data and {\tt GENBB}, the data in each individual mode are binned in 0.5~GeV$^2$-wide $\qsq$ bins. The signal yields after the final $|U|<50$~MeV requirement is compared between fit configurations with the background lineshape parameters allowed to vary up to $30\%$ from {\tt GENBB} (chosen to be large, without any loss of generality), with the background lineshapes fixed to {\tt GENBB}. To accumulate larger sample sizes, the check is repeated after integrating over $\qsq$. In both instances, no significant deviations in the yields are found because of the background lineshape variation. The variable $U$ represents the resolution in the reconstructed missing neutrino energy. Therefore, the difference between data and simulation in the signal lineshape is driven by differences in the resolution, accounted for by the $\kappa=5\%$ choice. As a check, $\qsq$-binned fits are performed and the signal yields are compared, allowing for a $5\%$ difference in resolutions between data and {\tt GENBB}; no systematic bias is seen due to this variation. As a conservative choice, the difference in results between  $\kappa=0\%$ and $\kappa=5\%$ is assigned as a systematic uncertainty.

Table~\ref{table:final_results} lists the baseline CLN and ($N=2$) BGL results including Gaussian constraints to the Belle-16~\cite{Glattauer:2015teq} data.

\begin{table}
   \caption{\label{table:final_results} The nominal $N=2$ results including systematic uncertainties. The normalizations for $|\Vcb|$ are from the $\deriv \Gamma /\deriv \qsq$ data in Ref.~\cite{Glattauer:2015teq}.}
  \begin{center}
     \begin{tabular}{cr|cr} \\ \hline \hline
      BGL $N=2$  & value\phantom{dum}                      & CLN & value\phantom{dum}\\ \hline
      $|\Vcb|\times 10^3$ & \!\!$ 41.09\pm1.16 $  & $|\Vcb|\times 10^3$  & \!\!$40.90\pm 1.14$\\
      $a^{f_+}_0\times 10$ & $ 0.126\pm 0.001$  & $\mathcal{G}(1)$  & $1.056\pm 0.008$  \\
      $a^{f_+}_1$ & $ -0.096\pm 0.003$  & $\rho^2_D$  & $1.155\pm0.023$  \\
      $a^{f_+}_2$ & $ 0.352\pm 0.053$    &   &  \\
      $a^{f_0}_1$ & $ -0.059\pm 0.003$   &   &  \\
      $a^{f_0}_2$ & $ 0.155\pm 0.049$    &   &  \\
     \hline \hline
    \end{tabular}
   \end{center}
\end{table}

\section{Discussion}

\subsection{Alternative determination of \boldmath$|\Vcb|$ using HFLAV branching fractions}

The differential rate given by Eq.~\ref{eqn:rate_formula} is integrated over $\qsq$ and $\ctl$ to obtain the total decay rate $\Gamma$. This is written in the form $\Gamma'= \Gamma/|V_{cb}|^2$ to strip the normalization off the $|\Vcb|$ component. Knowledge of the total branching fraction, $\mathcal{B}$, and the $B$ meson lifetime, $\tau$, allows the extraction of $|\Vcb|$ as
\begin{linenomath}
\begin{align} 
|\Vcb| = \sqrt{\frac{\mathcal{B}}{\Gamma'\tau_B}}.
\label{eqn:vcbfromgp}
\end{align} 
\end{linenomath}
The lifetimes are taken from HFLAV~\cite{HFLAV21} as $\tau_{B^+}=1.519 \pm 0.004$~ps and $\tau_{B^0}=1.638 \pm 0.004$~ps. 

The HFLAV~\cite{HFLAV21} values of the branching fractions used here are listed in Table~\ref{table:hflav21_data}. These numerical values are updated relative to those in the original articles~\cite{Aubert:2009ac,Glattauer:2015teq}, incorporating the latest available $D$ meson decay branching fractions. The resulting values of $|\Vcb|$, extracted from HFLAV and using Eq.~\ref{eqn:vcbfromgp}, are listed in Table~\ref{table:hflav21_data}.

\begin{figure*}
\centering
\includegraphics[width=3.4in]{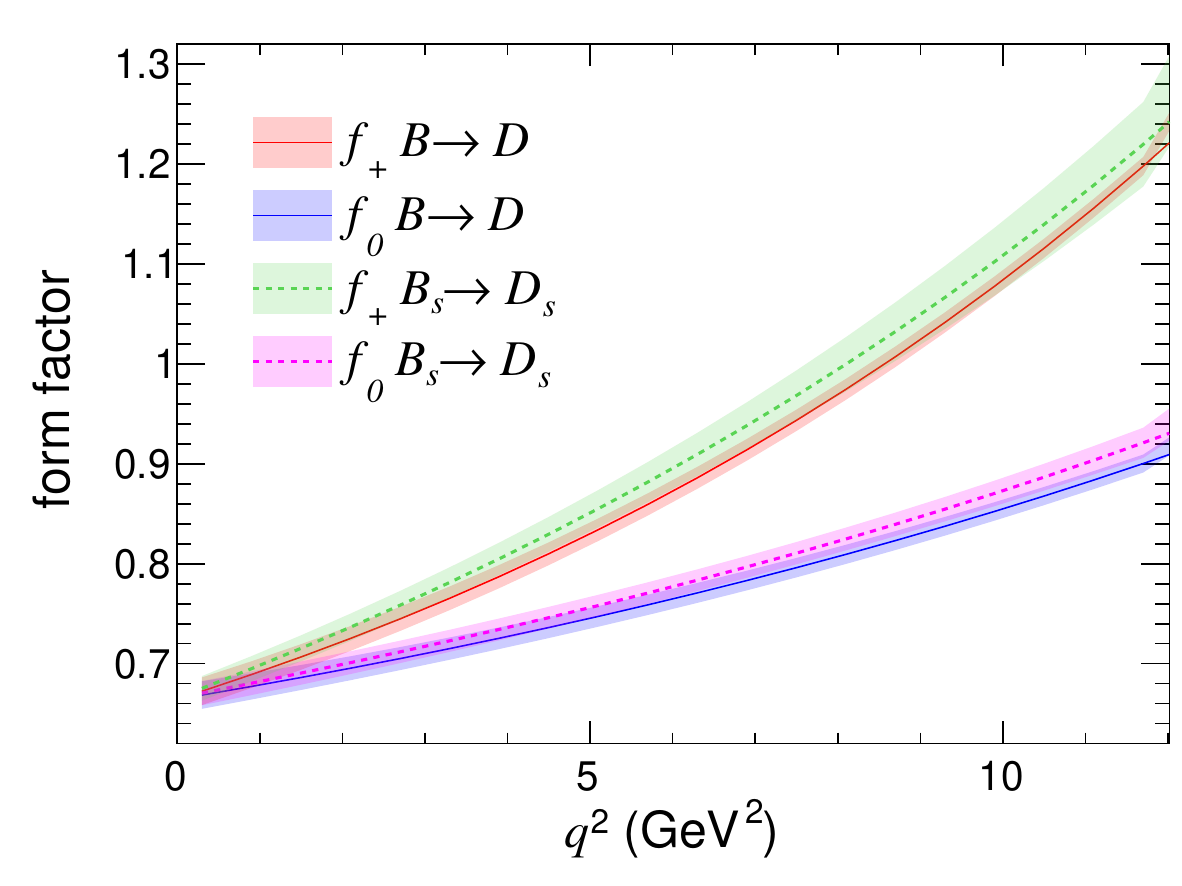}
\includegraphics[width=3.4in]{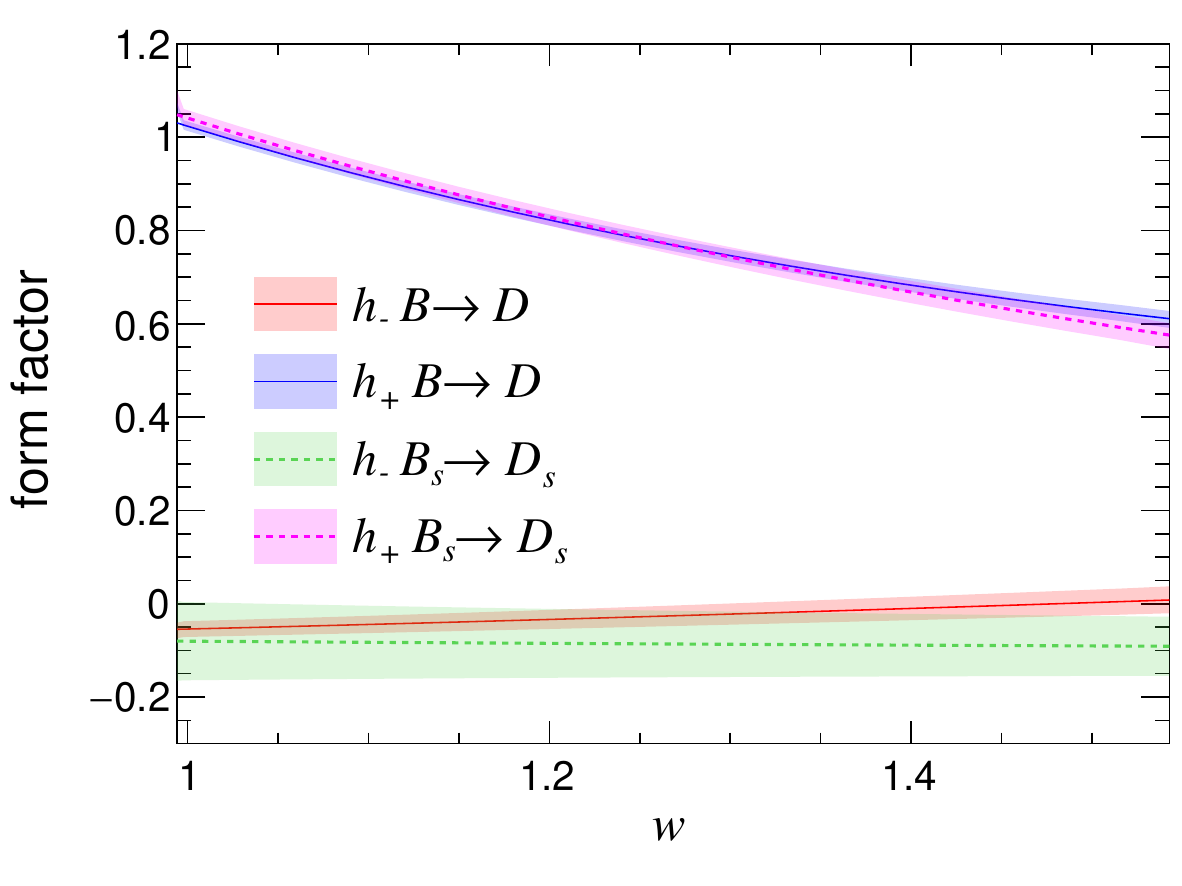}
\caption[]{$B_{(s)}\to D_{(s)}$ FFs as functions of $\qsq$ and $w$ using results from Table~\ref{table:final_results} and Ref.~\cite{McLean:2019qcx}: (left) $f_{+,0}$ and (right) $h_{\pm}$ bases. The filled areas correspond to $\pm1\sigma$ uncertainty envelopes.}
\label{fig:ff_comparisons}
\end{figure*}

\begin{table}[h]
   \caption{ \label{table:hflav21_data} Re-weighted $\bdlnu$ branching fractions as listed in HFLAV~\cite{HFLAV21} and the corresponding $|\Vcb|$ values extracted using the $\Gamma'$ (BGL) obtained from Table~\ref{table:final_results}.}
  \begin{center}
     \begin{tabular}{l c c} \\ \hline \hline
     Measurement & $\mathcal{B}(\bdlnu) \times 10^2$ & $|\Vcb|\times 10^3$ \\ \hline
     \babar-10~\cite{Aubert:2009ac} & $\mathcal{B}_{\Bz} = (2.15 \pm 0.11 \pm 0.14)$ & $ 40.02\pm1.76$ \\ 
     \babar-10~\cite{Aubert:2009ac} & $\mathcal{B}_{B^+} = (2.16 \pm 0.08 \pm 0.13)$ &  $38.67\pm1.41$  \\ 
     Belle-16~\cite{Glattauer:2015teq} & $\mathcal{B}_{\Bz} = (2.33 \pm 0.04 \pm 0.11)$ & $41.66\pm1.22$  \\ 
     Belle-16~\cite{Glattauer:2015teq} & $\mathcal{B}_{B^+} = ( 2.46 \pm 0.04 \pm 0.12)$ &  $41.27\pm1.23$ \\ \hline \hline
    \end{tabular}
   \end{center}
\end{table}

The values of $|\Vcb|$ extracted using exclusive $\bdlnu$, shown in Tables~\ref{table:final_results} and~\ref{table:hflav21_data} tend to be higher than $|\Vcb|=(38.36\pm0.90)\times10^{-3}$ obtained from exclusive $\bdstlnu$~\cite{Dey:2019bgc}. The last two values in Table~\ref{table:hflav21_data}, drawn from Belle-16~\cite{Glattauer:2015teq}, are the largest. Given the spreads (but compatible within quoted uncertainties) in the $|\Vcb|$ values from $\bdlnu$ between the CLN and BGL parameterizations (Table~\ref{table:final_results}) and the different tag-side normalization methods (Table~\ref{table:hflav21_data}), it is difficult to draw a clear conclusion. This is slightly different from the $\bdstlnu$~\cite{Dey:2019bgc} case, where a more robust value of $|\Vcb|$ was generally found. It is to be noted that a preliminary Belle~II untagged result~\cite{Belle-II:2022ffa} reports $|\Vcb|=(38.28\pm 1.16)\times10^{-3}$ from $\bdlnu$, more consistent with $|\Vcb|$ from $\bdstlnu$.

The $|\Vcb|$ values from $\bdlnu$ in Table~\ref{table:final_results} are higher than those typically obtained in $\bdstlnu$ and are closer to the inclusive value of $|\Vcb|=(42.16\pm 0.51)\times10^{-3}$~\cite{Bordone:2021oof}.

\subsection{SM prediction for \boldmath$\mathcal{R}(D)$}

Employing the definition of $\mathcal{R}(D)$ from Sec.~\ref{sec:rd_def} and the results presented in Table~\ref{table:final_results}, the SM prediction from this analysis (BGL) is
\begin{align}
\mathcal{R}(D)\Big|^{\scriptsize \babar}_{\rm SM\;theory} = 0.300 \pm 0.004.
\end{align}
This is consistent with other theoretical calculations and is compatible with the summer-2023 experimental measurement average~\cite{HFLAV21} of $0.357\pm0.029$ at 1.97 standard deviations.

\subsection{Comparisons with \boldmath$B_s\to D_s$ FFs}

Recently, the HPQCD Collaboration has published~\cite{McLean:2019qcx} FFs for $B_s\to D_s$ over the entire $\qsq$ range using the so-called heavy-HISQ action. Figure~\ref{fig:ff_comparisons} shows the comparisons in the two sets of FF bases described in Eq.~\ref{eqn:rel_hqet_conv_ff}. In the HQET limit at $\qsq\to q^2_{\rm max}$, $h_+\to +1$ and $h_-\to 0$. Assuming SU(3) symmetry among the three lightest quarks, the two sets of FFs should be equivalent. However, quark SU(3) symmetry is not a perfect symmetry. 

The extracted $B\to D$ form factors have better precision but show overall good agreement with the full-$\qsq$ $B_s\to D_s$ HPQCD Collaboration calculation, assuming flavor SU(3) symmetry. Some slight tension is visible in the HQET basis, at the maximum recoil point, $\qsq \to 0$, but otherwise flavor SU(3) symmetry seems to hold in the  $B_{(s)}\to D_{(s)}$ sector, consistent with the HQET analysis in Ref.~\cite{Bordone:2019guc}. These observations have implications for SU(3) flavor symmetry applicable to the $B_{(s)}\to D^\ast_{(s)}$ case, since a full-$\qsq$ HISQ calculation is already available~\cite{Harrison:2021tol}. One difference between the $B_{(s)}\to D_{(s)}$ and $B_{(s)}\to D^\ast_{(s)}$ cases is that for the former there are only two form factors that are strongly correlated at $\qsq = 0$ by the relation in Eq.~\ref{eqn:maxrecoilrel}. While a similar kinematic relation exists for the $B_{(s)}\to D^\ast_{(s)}$ case between the axial form factors, there are three axial and one vector FF; therefore the situation is much less constrained. The $\qsq \to 0$ relations are important in the HISQ formulation, to be able to perform the extrapolation to the physical quark masses~\cite{Harrison:2021tol}.

The comparisons in Fig.~\ref{fig:ff_comparisons} demonstrate that the role of the spectator quark, and therefore SU(3) symmetry breaking, cannot be very large. It has yet to be seen if similar relations hold between the FFs for $B\to D^\ast$ and $B_s\to D_s^\ast$, as expected in HQET~\cite{Bordone:2019guc}.

\section{Synthetic data}

\begin{figure}
\centering
\includegraphics[width=3.4in]{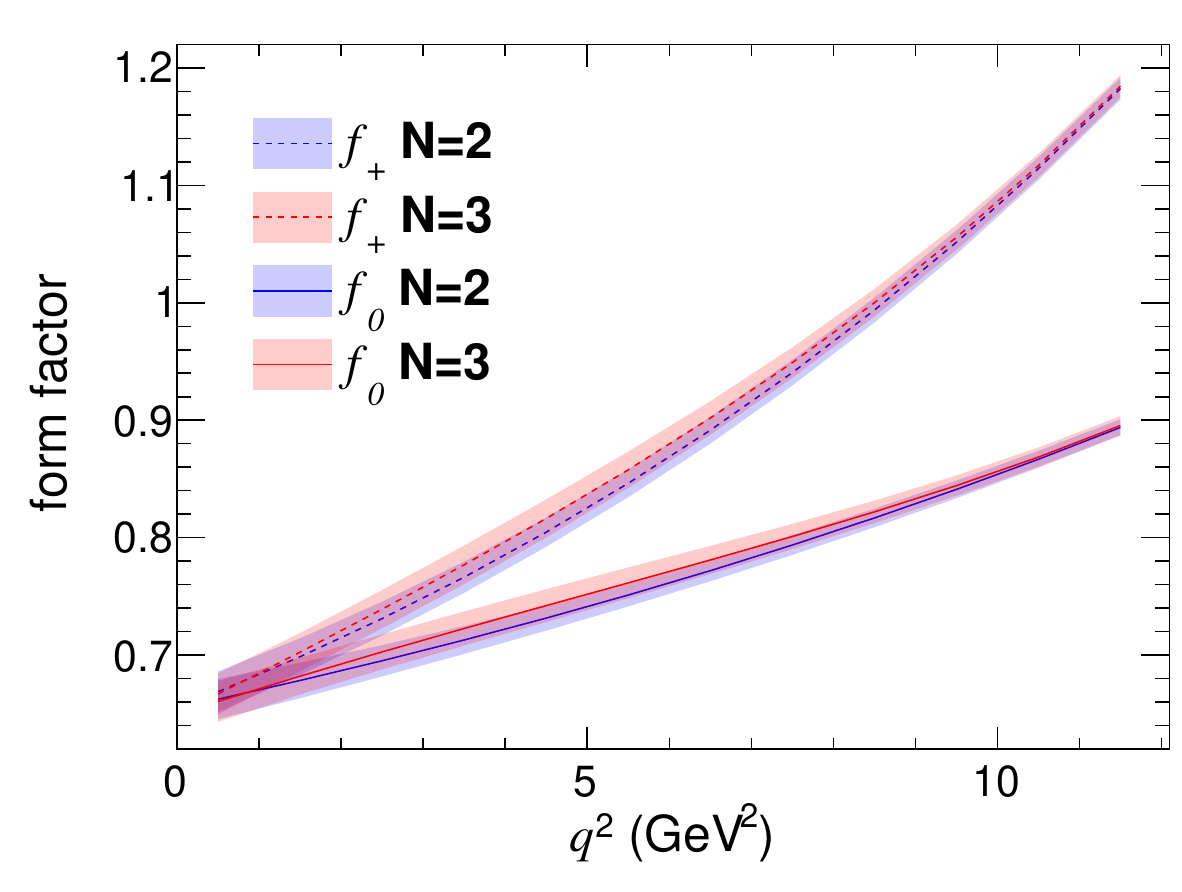}
\caption[]{Comparisons between the $N=2$ and $N=3$ BGL fit results using the \babar\ and FNAL/MILC data. The $1\sigma$ error envelopes include both the statistical and systematic uncertainties.}
\label{fig:ff_comparisons_N2_N3}
\end{figure}

In the fit method described in Sec.~\ref{sec:uml_fits}, the acceptance correction that depends on the form factor model is executed via the normalization integrals. The unfolded kinematic distributions are subsequently obtained from the resultant fit model after the minimization procedure. As long as the form factor parametrization has enough freedom, the fit results including the covariance matrix are fully representative of the statistical information in the data. The BGL $z$-expansion can be taken as a generic expansion, ignoring the physics interpretations imposed via the unitarity constraints. As mentioned in Sec.~\ref{sec:bgl_results}, the $-2\ln \mathcal{L}(\vec{x})\LARGE|_{\tiny \babar}$ component of the minimization function is unchanged at the optimal points, between the $N=2$ and $N=3$ BGL fits. From Table~\ref{table:final_results}, the $N=2$ results are consistent with unitarity and are the nominal results. The $N=3$ results in Table~\ref{table:bgl_results_N3} violate unitarity, but can still be taken as a generic expansion.  Figure~\ref{fig:ff_comparisons_N2_N3} further demonstrates the consistency between the $N=2$ and $N=3$ fits.

The statistical uncertainties provided by the covariance matrices assume parabolic uncertainties around the minimization points. This is validated by checking that the uncertainties provided via the {\tt MINOS} routine are symmetric, both at $1\sigma$ and $2\sigma$. The {\tt MINOS} uncertainties are always found to agree with those from the {\tt HESSE} routine. The numeric data are provided in the file {\tt BaBar\_Dlnu\_2023\_BGL\_results.h} and the exact BGL form to be used is provided in the file {\tt B2D\_BGL.h}. To ascertain the effect of the uncertainties in the FNAL/MILC calculations, the central values of the lattice data are smeared according to the corresponding covariance matrix and the \babar+lattice BGL fits are repeated for $10^6$ instances. The spread in the fit results is employed to estimate the covariance matrix for the lattice contribution to total the uncertainties, $C_{\rm lat}$. The uncertainties solely due to the {\babar} data can then be estimated as $C_{\scriptsize \babar} = C_{\rm tot} - C_{\rm lat}$, where $C_{\rm tot}$ is the nominal uncertainty from the fit results with the lattice data information incorporated via Gaussian constraints. The numeric results for $ C_{\rm lat}$ are provided in the aforementioned file. As an example, the decomposition of the uncertainty for the $N=2$ BGL fit including {\babar} and lattice data, is given in Table~\ref{table:hflav21_data_stat}.

\begin{table}[h]
   \caption{\label{table:hflav21_data_stat} Re-weighted $\bdlnu$ branching fractions as listed in HFLAV~\cite{HFLAV21} and the corresponding $|\Vcb|$ values extracted using the $\Gamma'$ ($N=2$, BGL) obtained from the \babar-1 fits in Table~\ref{table:bgl_results_N2}. The quoted uncertainties correspond to \babar\;(present analysis), FNAL/MILC and HFLAV, respectively.}
  \begin{center}
     \begin{tabular}{l c} \\ \hline \hline
     $\mathcal{B}$ measurement & $|\Vcb|\times 10^3$ \\ \hline
     \babar-10~\cite{Aubert:2009ac}    &  $40.36 \pm 0.17 \pm 0.10 \pm 1.67$ \\ 
     \babar-10~\cite{Aubert:2009ac}    &  $38.98 \pm 0.15 \pm 0.09 \pm 1.30$  \\ 
     Belle-16~\cite{Glattauer:2015teq} &  $42.01 \pm 0.18 \pm 0.10 \pm 1.06$  \\ 
     Belle-16~\cite{Glattauer:2015teq} &  $41.60 \pm 0.17 \pm 0.10 \pm 1.07$ \\ \hline \hline
    \end{tabular}
   \end{center}
\end{table}

To facilitate using the results from this article, synthetic data are generated for $f_{+,0}$ at 12 equidistant $\qsq$ points from 0.5 to 11.5~GeV$^2$, resulting in 24 synthetic data points, for each of the fit configurations. The numeric data are provided in the file {\tt BaBar\_Dlnu\_2023\_BGL\_synthdata.h}. The 24 data points are however not independent and a judicious subset of 5(7) data points for the $N=2(3)$ BGL fit configurations should be taken, in line with the number of free fit parameters in the $z$-expansion, so that the corresponding reduced covariance matrix is invertible.

\section{Summary}

In summary, the first two-dimensional unbinned angular analysis in $\{\qsq, \ctl\}$ for the process $\bdlnu$ is reported. A novel event-wise signal-background separation technique is utilized that preserves multidimensional correlations present in the data. The angular fit incorporates acceptance correction in the $\{\qsq, \ctl\}$ phase space and accounts for different $\bdlnu$ reconstruction modes having independent characteristics. It is shown that within statistical precision, the lepton helicity distribution follows a $\sin^2 \thetal$ distribution, as expected in the Standard Model. This bolsters confidence in the hadronic tagging procedure, acceptance correction, and signal-background separation techniques. In the future, model-independent new-physics contributions can be probed via searches for additional $\ctl$ terms in the semileptonic sector, as has already been studied in the electroweak penguin sector at LHCb~\cite{LHCb:2014auh}.

High-precision $N=\{2,3\}$ BGL fits to the $B\to D$ form factors are reported and found to be consistent with $B_s\to D_s$ form factors from lattice, as expected from quark SU(3) relations. The form factors give the SM prediction $\mathcal{R}(D) = 0.300 \pm 0.004$. Combined with the differential branching fraction data from Belle~\cite{Glattauer:2015teq}, the BGL results yield the result for $|\Vcb|$ from exclusive $B\to D$ as $ 41.09\pm1.16$, closer to the inclusive value, than $|\Vcb|$ from exclusive $B\to D^\ast$.

\begin{acknowledgments}
\section{Acknowledgements}

We are grateful for the extraordinary contributions of our PEP-II colleagues in achieving the excellent luminosity and machine conditions that have made this work possible. The success of this project also relies critically on the expertise and dedication of the computing organizations that support BABAR, including GridKa, UVic HEP-RC, CC-IN2P3, and CERN. The collaborating institutions wish to thank SLAC for its support and the kind hospitality extended to them. We also wish to acknowledge the important contributions of J.~Dorfan and our deceased colleagues E.~Gabathuler, W.~Innes, D.W.G.S.~Leith, A.~Onuchin, G.~Piredda, and R. F.~Schwitters.

\end{acknowledgments}

\addcontentsline{toc}{section}{References}
\setboolean{inbibliography}{true}
\bibliography{main}

\end{document}